%% file: ATLASHIP14.tex
\documentclass[cernpreprint,UKenglish,texlive=2011,txfonts,texmf]{atlasdoc}
\pdfoutput=1
\usepackage[biblatex=false]{atlaspackage}
\usepackage{atlascontribute}
\usepackage{atlasphysics}
\usepackage{cite}

\AtlasTitle{Search for magnetic monopoles and stable particles with high electric charges in 8~TeV $pp$ collisions with the ATLAS detector}

\author{The ATLAS Collaboration}

\PreprintIdNumber{CERN-PH-EP-2015-174}

\AtlasDate{\today}

\arXivId{1509.08059}

\AtlasJournalRef{Phys. Rev. D 93, 052009 (2016)}
\AtlasDOI{10.1103/PhysRevD.93.052009}

\AtlasAbstract{%
  A search for highly ionizing particles produced in proton-proton
  collisions at 8~TeV center-of-mass energy is performed by the ATLAS
  collaboration at the CERN Large Hadron Collider. The dataset used
  corresponds to an integrated luminosity of 7.0~fb$^{-1}$. A customized
  trigger significantly increases the sensitivity,
  permitting a search for such particles with charges and
  energies beyond what was previously accessible.  No events were found
  in the signal region, leading to production cross section
  upper limits in the mass range 200--2500~GeV for magnetic monopoles with
  magnetic charge in the range $0.5g_{\rm D}<|g|<2.0g_{\rm D}$, where
  $g_{\rm D}$ is the Dirac charge, and for stable particles with
  electric charge in the range $10<|z|<60$. Model-dependent limits are
  presented in given pair-production scenarios, and model-independent
  limits are presented in fiducial regions of particle energy and
  pseudorapidity.  }

\begin{document}

\maketitle

\section{Introduction}
\label{sec:intro}

The multi-TeV energy regime accessible at the CERN Large Hadron
Collider (LHC) enables the exploration of uncharted territories of particle physics. A new massive particle would represent a dramatic deviation from the predictions of the Standard Model, and such a spectacular discovery would lead to fundamental insights and critical theoretical developments. This paper presents a dedicated search for a long-lived highly ionizing particle (HIP) signature in the ATLAS detector. Such a signature differs from those of the known objects (e.g., electrons, muons, and jets) reconstructed in ATLAS and would be missed by analyses that rely only on such objects. 
HIP signatures can arise at LHC energies as an important feature of physics beyond the Standard Model, for example, in theories of magnetic monopoles and dyons, strange quark matter, $Q$-balls, and stable microscopic black-hole remnants~\cite{Fairbairn2007,MoEDAL2014}.

The Dirac argument~\cite{Dirac1931,Dirac1948} addresses the problem of electric charge quantization by postulating the existence of particles possessing magnetic charge. The lightest magnetic monopole would be stable and carry a magnetic charge that is a multiple of the Dirac charge $g_{\rm D}$, i.e., in Gaussian units,  

\begin{equation}
\frac{g_{\rm D}e}{\hbar c}=\frac{1}{2}\hspace{0.3cm}
\Rightarrow \hspace{0.3cm} \frac{g_{\rm
    D}}{e}=\frac{1}{2\alpha_{\rm e}}\approx 68.5,
\label{eqn:gcharge}
\end{equation}
where $e$ is the elementary electric charge and $\alpha_{\rm e}$ is the fine
structure constant. With the introduction of a magnetic monopole, the
duality of Maxwell's equations implies a magnetic coupling

\begin{equation}
\alpha_{\rm m} = \frac{g_{\rm D}^2}{\hbar c} = \frac{1}{4\alpha_{\rm e}}, \\
\label{eqn:magcoupl}
\end{equation}
which is very large, precluding any perturbative calculation of monopole production processes. In terms of ionization energy loss at high velocity, a monopole with the Dirac charge corresponds to an electrically charged particle with charge \mbox{$|z|\approx 68.5$}. A monopole would thus manifest itself as a HIP, as would any highly charged stable particle. In addition to the Dirac argument, topological monopole solutions arise naturally in unification theories with gauge symmetry breaking~\cite{Polyakov1974,tHooft1974}. Monopole solutions are also allowed in the electroweak theory itself with a mass at the TeV scale and an elementary magnetic charge that is twice the Dirac charge~\cite{Cho1997,MoEDAL2014}. 

Searches for monopoles have been carried out in cosmic-ray experiments~\cite{MACRO2002a,SLIM2008a,RICE2008,ANITA2011,ANTARES2012,IceCube2013,IceCube2014}, in matter~\cite{Ross1973,Kovalik1986,Jeon1995,Bendtz2013a}, and at colliders~\cite{Aubert:1982zi,Musset:1983ii,D01998,Aktas:2004qd,CDF2006,Fairbairn2007,OPAL2008,ATLAS2012a,DeRoeck2012b,Bendtz2014}. The high luminosity and energy of LHC collisions mean that monopoles (and other HIPs) can be probed at higher masses and to greater precision than was previously accessible~\cite{DeRoeck2012a}. 
In 2010, ATLAS initiated the search for HIPs at the LHC by considering a particle producing a region of high ionization density in the transition radiation tracker (TRT) and slowing down and stopping in the electromagnetic (EM) calorimeter~\cite{ATLAS2011a}. Since energy loss by bremsstrahlung and $e^+e^-$ pair production is negligible for HIPs, the ionization energy deposit in the EM calorimeter is narrower than that associated with electrons and photons, which induce an EM shower. This stopping signature applies to HIPs with charge \mbox{$|z|\gtrsim10$}, while particles with lower charges have been probed at ATLAS and CMS using a muon-like signature~\cite{ATLAS2015,CMS2013a}. The stopping signature was used at ATLAS to set the first constraints on the production of magnetic monopoles carrying a single Dirac charge \mbox{($|g|=1.0g_{\rm D}$)} in $pp$ collisions at 7~TeV center-of-mass energy~\cite{ATLAS2012a}.  This first monopole search at the LHC relied on an electron trigger.  A new dedicated ATLAS trigger designed to improve the sensitivity to the stopping HIP signature and access new regions of HIP charge is used in the present search. Further improvements with respect to the previous analyses include a larger integrated luminosity, higher center-of-mass energy, extension of the signal acceptance to the detector forward regions (pseudorapidity\footnote{ATLAS uses a right-handed coordinate system with its origin at the nominal interaction point (IP) in the center of the detector and the $z$-axis along the beam pipe. The $x$-axis points from the IP to the center of the LHC ring, and the $y$-axis points upward. Cylindrical coordinates $(r,\phi)$ are used in the transverse plane, $\phi$ being the azimuthal angle around the beam pipe. The pseudorapidity is defined in terms of the polar angle $\theta$ as \mbox{$\eta=-\ln\tan(\theta/2)$}.} up to $|\eta|=2$), interpretation for a magnetic charge $|g|$ up to twice the Dirac charge as well as for an electric charge $|z|$ between 20 and 60, and an interpretation for spin-0 HIPs in addition to spin-1/2 for the model-dependent limits.

\section{ATLAS detector}
\label{sec:detector}

The ATLAS experiment~\cite{ATLAS2008} is a multipurpose particle physics detector with a forward-backward symmetric cylindrical geometry and near 4$\pi$ coverage in solid angle. In the ATLAS detector, the HIP signature can be readily distinguished using the transition radiation tracker in the inner detector (ID) and the liquid-argon sampling electromagnetic calorimeter.

Tracking in the inner detector is performed by silicon-based detectors and an outer tracker, the TRT, using straw tubes with particle identification capabilities based on transition radiation. The TRT is divided into barrel (covering the pseudorapidity range \mbox{$|\eta|<1.0$}) and endcap \mbox{($0.77<|\eta|<2.0$)} components. A track typically comprises 32 straw hits. In the front-end electronics of the TRT, discriminators are used to compare the straw-tube signal against low and high thresholds. 
HIPs would produce a large number of high-threshold (HT) hits along their trajectories, due to both the high ionization of the HIP and the high density of $\delta$-rays emitted from the material along the trajectory of the HIP. The amount of ionization in a straw tube needed for a TRT HT hit is roughly equivalent to three times that expected from a minimum ionizing particle.

A thin superconducting solenoid magnet surrounding the tracking section of the ATLAS detector produces a field of approximately 2~T parallel to the beam axis. The ID and solenoid together represent an amount of material of approximately two radiation lengths for \mbox{$|\eta|<0.7$} and three radiation lengths elsewhere.  

Liquid-argon sampling EM calorimeters, which comprise accordion-shaped electrodes and lead absorbers, surround the ID and solenoid. The EM calorimeter in the pseudorapidity ranges \mbox{$|\eta|<1.475$} (barrel) and \mbox{$1.375<|\eta|<2.5$} (endcap) is segmented transversely and divided into three layers in depth,  denoted first (EM1), second (EM2), and third (EM3) layer, respectively. In the pseudorapidity range \mbox{$|\eta|<1.8$}, an additional presampler layer in front of the accordion calorimeter is used to provide a measurement of the energy lost in front of the calorimeters~\cite{ATLAS2008}. The presampler, EM1, and EM2 layers in the barrel represent an amount of material of approximately 0.5, 4.3, and 16.5 radiation lengths, respectively. The noise level in the EM calorimeter is typically 200~MeV or less.
The robustness of the EM calorimeter energy reconstruction has been studied in detail and pulse shape predictions are consistent with the measured signals~\cite{ATLAS2010b}.

Beyond the EM calorimeter, in the barrel region, the ATLAS hadronic calorimeter is made of scintillator tiles and steel absorber plates. It comprises a barrel in the pseudorapidity range \mbox{$|\eta|<1.0$} and an extended barrel in the range \mbox{$0.8<|\eta|<1.7$}. Liquid-argon hadronic endcap calorimeters cover the range \mbox{$1.5<|\eta|<3.2$}. The noise level in the hadronic calorimeter is typically 100~MeV or less.

The ATLAS data were filtered by a three-level trigger system that reduced the rate from 20~MHz to $\sim$400~Hz.  Level-1 is a hardware-based trigger that, for the purposes herein, identifies regions of interest (RoI) associated with energy deposits in the calorimeter. Level-2 and Event Filter triggers are implemented in software, with detector information corresponding to the RoI accessible by the Level-2 trigger, whereas the full detector information is accessible by the Event Filter.

The stopping power of a HIP in matter depends on its charge, mass, and energy (but not on its spin), as well as the material traversed along its path. 
Details of the ATLAS geometry are given in Ref.~\cite{ATLAS2008} in terms of number of radiation lengths $X_{\rm 0}$, as a function of depth and pseudorapidity. In this search, a HIP candidate must deposit energy in the EM calorimeter to be selected by the Level-1 trigger. In 8~TeV collisions, this limits the range of HIP charges that can be probed in ATLAS to \mbox{$|g|\le2.0 g_{\rm D}$} for magnetic charge and \mbox{$|z|\le60$} for electric charge.

\section{Simulations}
\label{sec:simulations}

The {\sc MadGraph5} Monte Carlo (MC) event generator~\cite{MadGraph} is used to estimate production cross sections and to generate signal events where HIPs are produced in pairs from the initial $pp$ state via quark-antiquark annihilation into a virtual photon. This process is modeled by assuming leading-order Drell-Yan (DY) heavy charged-particle pair production, where the coupling is obtained by scaling the photon-electron coupling by the square of the HIP electric or magnetic charge (e.g., a factor 68.5$^2$ for a Dirac monopole). In the absence of a consistent theory describing the coupling of the HIP to the $Z$ boson, such a coupling is set to zero in the {\sc MadGraph5} model. 
HIP production models suffer from large uncertainties due to the large coupling of the HIP to the photon precluding any perturbative calculation beyond leading order. For magnetic monopole pair production, the coupling is described by Eq.~(\ref{eqn:magcoupl}). 
The CTEQ6L1~\cite{CTEQ6L1} parton distribution functions of the proton are employed and \PYTHIA version 8.175~\cite{Pythia6,Pythia8} is used for the hadronization and the underlying-event generation. Direct pair production implies that the HIPs are not part of a jet and are thus isolated.

Given the production model uncertainties, the impact that a change in model would have on the angular distributions and cross sections is investigated by also considering spin-0 DY HIP production. In addition to lower cross sections, angular momentum conservation dictates that DY production of spin-0 HIPs is suppressed near the phase-space thresholds due to the fact that the intermediate (virtual) photon has spin-1. Thus, spin-1/2 and spin-0 HIPs have different angular distributions, providing a measure of how model uncertainties affect the search acceptance.  The spin-0 samples are generated using {\sc MadGraph5}, as described above.  

The model-independent interpretation does not assume a particular production mechanism. For this, single-particle HIP samples with uniform distributions in HIP kinetic energy and pseudorapidity, in the ranges \mbox{$E_{\rm kin}<3000$~GeV} and \mbox{$|\eta|<2.5$}, respectively, are used to determine the selection efficiencies in regions of kinematic phase space. Since the interaction of HIPs with material is
spin-independent~\cite{Ahlen1975,Ahlen1978}, these efficiencies are identical for spin-0 and spin-1/2 HIPs.

The DY and single-particle samples, which have approximately 20,000 and 50,000 events, respectively, are produced for HIPs with masses $m$ equal to 200, 500, 1000, 1500, 2000, and 2500~GeV. For each mass point, magnetic monopoles are simulated for magnetic charges $|g|$ (in units of the Dirac charge $g_{\rm D}$) 0.5, 1.0, 1.5, and 2.0. Separate samples of HIPs are produced with electric charges $|z|$ (in units of the elementary charge) $10$, $20$, $40$, and $60$.

The single-particle and spin-1/2 DY samples are processed by the ATLAS detector simulation~\cite{ATLAS2010c} based on {\sc GEANT4}~\cite{GEANT42003}. In addition to the standard ionization process based on the Bethe-Bloch formula, the particle interaction model includes secondary ionization by $\delta$-rays.  For monopoles, a modified Bethe-Bloch formula is used to account for the velocity-dependent Lorentz force~\cite{Ahlen1975,Ahlen1978}. The effect of the ATLAS solenoid magnetic field (bending of trajectories of electrically charged particles and acceleration of magnetic monopoles) is included in the equations of motion. A correction for electron-ion recombination effects in the EM calorimeter (Birks' law) is applied, with typical visible energy fractions between 0.1 and 0.4 for the signal particles considered~\cite{Burdin2012}. 
Trigger efficiency losses for slow particles arriving at the calorimeter later than highly relativistic particles (and therefore being assigned to the wrong bunch crossing) are simulated.
Particles arising from multiple interactions in the same or neighboring bunch crossings (``pileup'') are overlaid on both the pair-production and single-particle samples to reflect the conditions of the data sample considered in the search. This full detector simulation of HIPs uses significant computing resources and, hence, was not performed for spin-0 DY HIPs. 

A data-driven method is used to estimate backgrounds surviving the
final selections (see Section~\ref{sec:backgrounds}). Two samples of
simulated background events are used to increase confidence in the
modeling of the relevant observables. These are labeled
$W^{\pm}\rightarrow \nu e^{\pm}$ and ${\rm DY}\rightarrow e^+ e^-$ and
correspond to electroweak processes in which $W$ bosons, and $Z$ bosons or
virtual photons, decay to electrons. Both samples are generated with
\POWHEG~\cite{POWHEG} and then passed through \PYTHIA~8 with the AU2~CT10 set of tuned MC parameters ~\cite{ATLAS2012f} for hadronization and parton showering. 

\section{Trigger}
\label{sec:trigger}

At Level-2, standard ATLAS EM triggers implicitly require energy deposition in the EM2 layer and thus are unable to capture HIPs that stop in EM1 or in the presampler. Furthermore, conditions for 8~TeV collisions include either high thresholds on the transverse energy, \mbox{$E_{\rm T}=E\sin\theta$}, for photon triggers or tight requirements on track quality and isolation for electron triggers (severely impairing HIP searches due to the effects of long-range $\delta$-rays). Thus, a new Level-2 trigger dedicated to HIP searches was developed and deployed in 2012. The Level-2 HIP trigger has no EM energy requirements beyond Level-1 and yields the maximum acceptance to HIPs that the ATLAS geometry can possibly allow using calorimeter-based Level-1 triggers. Crucially, this provides access to HIPs with higher charges and lower energies. A low rate is achieved by imposing requirements on
the number and fraction of TRT HT hits in a narrow region around the Level-1 calorimeter RoI.

\subsection{HIP trigger selection}
\label{sec:trigger_definition}

The lowest threshold unprescaled Level-1 (L1) calorimeter trigger ~\cite{ATLAS2012d} in 2012 is used to seed the Level-2 HIP trigger. The L1 trigger selects calorimeter towers exceeding an $\eta$-dependent $E_{\rm T}$ threshold between 18 and 20~GeV and containing less than 1~GeV in the corresponding region of the hadronic calorimeter. The hadronic energy veto has a small impact on a HIP pair-produced signal in 8~TeV collisions, since only a negligible fraction of HIP candidates with equivalent charge $1.0g_{\rm D}$ or higher would possess enough energy to enter the hadronic calorimeter.

The HIP trigger algorithm reconstructs two variables: the number of TRT HT hits, $N_{\rm HT}^{\rm trig}$, and the fraction of all TRT hits that are HT hits, $f_{\rm HT}^{\rm trig}$, in a wedge of $\pm 0.015$~rad in $\phi$ defined within the Level-1 RoI. The center of this wedge is determined as the location of the bin with the highest number of TRT HT hits among 20 bins each of 0.01~rad in $\phi$ around the RoI center. The RoI $\eta$ information is also used to identify and count only the hits in the parts of the TRT that cover the corresponding $\eta$ regions. 

The selection was defined as $N_{\rm HT}^{\rm trig}>20$ and $f_{\rm HT}^{\rm trig}>0.37$ as a compromise between controlling the rate and ensuring a high signal efficiency. 
The rate of events passing these requirements is dominated by chance occurrences in multijet events where more HT hits than usual are produced in the $\phi$ wedge defined by the trigger, either due to overlapping charged particles within the same straws or due to electronic noise. 

\subsection{Trigger performance in 8 TeV collisions}
\label{sec:trigger_performance}

The HIP trigger rate was in the range 0.4--0.7~Hz from its deployment in September 2012 until December 2012. The integrated luminosity collected during this period was 7.0~fb$^{-1}$ of 8~TeV proton-proton collision data. 
The rate is found to be lower at higher instantaneous luminosities, which correspond to the beginning of the runs when more populated bunches produce higher pileup. This is explained by the observation that $f_{\rm HT}^{\rm trig}$ is sensitive to pileup: additional collisions per bunch crossing produce additional soft tracks that contaminate the $\phi$ wedge with low-threshold hits, thus reducing the HT hit fraction. This can affect the signal efficiency as well. 

The dedicated HIP trigger provides a considerable acceptance gain by capturing HIP candidates that stop in the first EM calorimeter layer, or even in the EM presampler. 
With 2012 pileup conditions, a monopole candidate that is within the acceptance of the TRT and has passed the Level-1 trigger requirements would have a high ($\gtrsim90\%$) probability to satisfy the HIP trigger algorithm. The efficiency drops off for HIP candidates of sufficiently high energy that have a high probability to penetrate through to the hadronic calorimeter and provoke the Level-1 hadronic veto.
The available models of HIP production predict the energy distribution to peak in the range 100--500~GeV (see Ref.~\cite{DeRoeck2012a} and references therein), in which a large fraction of $|g|=1.0g_{\rm D}$ monopole candidates are recovered by the HIP trigger, as compared to existing photon triggers. As an example, the HIP trigger acceptance times efficiency in the DY spin-1/2 monopole pair-production model for $|g|=1.0g_{\rm D}$ and $m=1000$~GeV is $(24.6\pm 0.3)$\%, while for the 120~GeV single-photon trigger it is only $(3.1\pm 0.1)$\%.  
For the charges and masses considered in this search, only HIPs with $\beta>0.4$ would be energetic enough to reach the EM calorimeter to be selected by the L1 trigger. The introduction of the HIP trigger reduces the minimum kinetic energy needed to trigger on $|g|=2.0g_{\rm D}$ monopoles from $\sim$1500~GeV to $\sim$900~GeV. 

\section{Event selection}
\label{sec:selection}

The event selection starts by identifying energy deposits (``clusters'') in the EM calorimeter and associating them with a region with a high fraction of HT hits in the TRT. EM cluster candidates are constructed by the EM topological cluster algorithm~\cite{Lampl2008}, which starts with a seed EM calorimeter cell with large signal-to-noise ratio, iteratively adds neighboring cells with a threshold defined as a function of the expected noise, and finishes by including all direct neighbor cells on the outer perimeter. This algorithm is very efficient for reconstructing clusters from HIP energy depositions. Topological cluster formation does not require energy deposits in EM2, allowing the reconstruction of clusters from HIPs that stop in EM1 or in the EM presampler in addition to those that stop in EM2. In the TRT barrel, the TRT hit-counting region is a rectangular {\it road} of constant width $\pm 4$~mm in the transverse plane centered around the region in $\phi$ with the highest density of HT hits. In the TRT endcap, a wedge of $\Delta\phi=\pm 0.006$ is used instead. The hit-counting procedure is described in more detail in Ref.~\cite{ATLAS2012a}. 

The selection is designed to reduce Standard Model backgrounds while retaining HIP signal candidates and relies on the following variables:

\begin{itemize}

\item $f_{\rm HT}$: the fraction of TRT HT hits in a road or wedge, as described above, matched to an EM cluster. Compared to how the TRT hits are counted in Ref.~\cite{ATLAS2012a}, a slight improvement is made in the central \mbox{($|\eta|<0.1$)} region and in the TRT barrel-endcap transition region \mbox{($0.77<|\eta|<1.06$)}, which yields a higher signal efficiency. In the central region, the TRT is split between $\eta<0$ and $\eta>0$ barrels and $f_{\rm HT}$ is computed separately for each TRT component. The maximum value obtained from either of these components separately or combined is selected as the new $f_{\rm HT}$ value. Similarly, in the transition region, $f_{\rm HT}$ is recomputed by considering the barrel and the endcap separately as well as together.

\item $E_{0}$, $E_{1}$, and $E_{2}$: the energy belonging to an EM calorimeter cluster contained in the presampler, EM1, and EM2, respectively.

\item $w_{\rm 0}$, $w_{\rm 1}$, and $w_{\rm 2}$: the fraction of EM cluster energy contained in the two most energetic cells in the presampler, four most energetic cells in EM1, and five most energetic cells in EM2, respectively. This provides a measure of the energy dispersions in each EM calorimeter layer, with values around unity (occasionally exceeding unity due to negative cell-noise energies) corresponding to the minimum dispersion, as expected for HIPs. The number of cells chosen was optimized by maximizing the discrimination power between HIPs and electron backgrounds, accounting for the different granularities in the different EM calorimeter layers.

\item $w$: a combination of the three energy dispersion variables above, defined as the arithmetic mean of all $w_i$ ($i$=0,1,2) for which $E_{i}$ exceeds a 5~GeV threshold. This threshold ensures that the energy dispersion in a layer that is not traversed by a HIP is not included, since this layer would mostly contain noise. 

\end{itemize}

The selection criteria, defined below, are chosen so as to minimally
impact the signal efficiency. The optimal $f_{\rm HT}$ and $w$ cut values that
define the signal region maximize the ratio of signal over square root
of the background across all mass and charge points. The background
contribution is obtained from $w$--$f_{\rm HT}$ pseudodata generated by
randomly sampling the individual one-dimensional distributions of
$f_{HT}$ and $w$ in collision data. In order to exclude the possibility of
generating data points from the signal region in the pseudodata, only
candidate events with $w < 0.8$ are used to generate the one-dimensional
$f_{\rm HT}$ distribution and candidate events with $f_{\rm HT} < 0.6$ are used to
generate the one-dimensional $w$ distribution.  
At each stage, events
without any candidates satisfying the criteria are discarded.

\begin{enumerate}

\item The HIP trigger criteria must be satisfied.

\item Preselection: clusters with \mbox{$E_{\rm T}>16$~GeV} in the EM calorimeter and associated with a region in the TRT satisfying \mbox{$f_{\rm HT}>0.4$} are selected. This efficiently identifies the cluster candidates that triggered the event, plus possible additional candidates in the same event. If multiple candidates are found within a window \mbox{$\Delta \phi \times \Delta \eta = 0.05 \times 0.1$}, only the cluster with the highest summed energy in the presampler and EM1 layers is kept.

\item EM layers: it is required that at least one of the \mbox{$E_{0}>5$~GeV} or \mbox{$E_{1}>5$~GeV} requirements is satisfied for the selected cluster candidate. This rejects backgrounds where there is only energy in EM2 (while a HIP penetrating EM2 must necessarily have also gone through the preceding layers).

\item Pseudorapidity: cluster candidates are selected in the range \mbox{$0<|\eta|<1.375$} or \mbox{$1.52<|\eta|<2.0$}. The EM calorimeter barrel-endcap transition regions are excluded to ensure the robustness of the $w$ variable.

\item Hadronic veto: cluster candidates with less than 1~GeV hadronic calorimeter energy calculated using the hadronic barrel and extended barrel calorimeters are selected. This criterion ensures that the efficiency of the Level-1 trigger hadronic veto is well accounted for in the simulation. 

\item Single candidate: in case of multiple candidates in the same event, only the candidate with highest $f_{\rm HT}$ is kept. This has a negligible impact on signal efficiencies while ensuring a consistent event-based background estimate from data. 

\item EM dispersion: candidates with \mbox{$w\geq0.94$} are selected.

\item TRT HT hits: candidates with \mbox{$f_{\rm HT}\geq0.70$} are selected.

\end{enumerate}

\begin{figure}[tb]
\begin{center}
\includegraphics[width=0.495\linewidth]{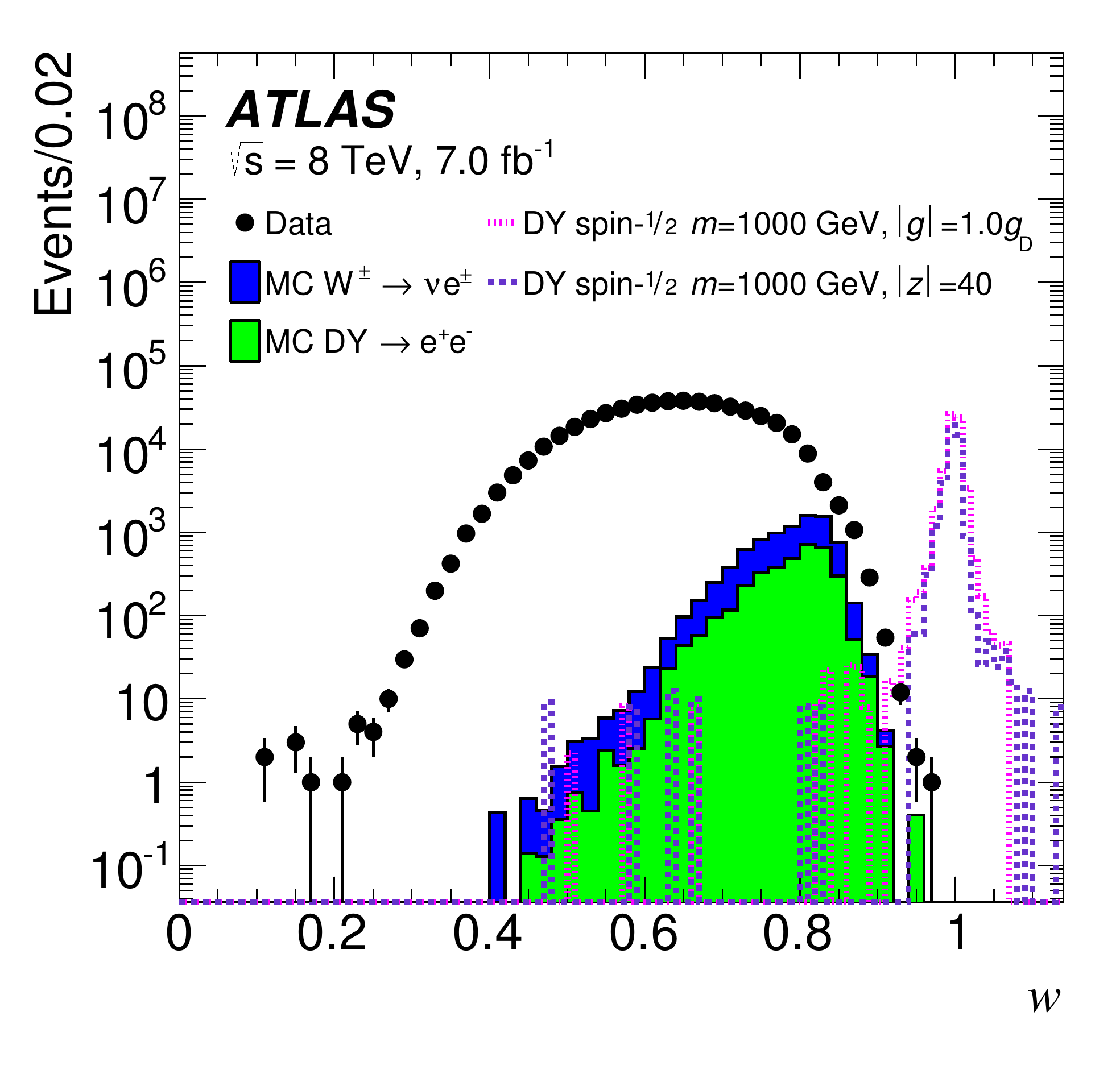}
\includegraphics[width=0.495\linewidth]{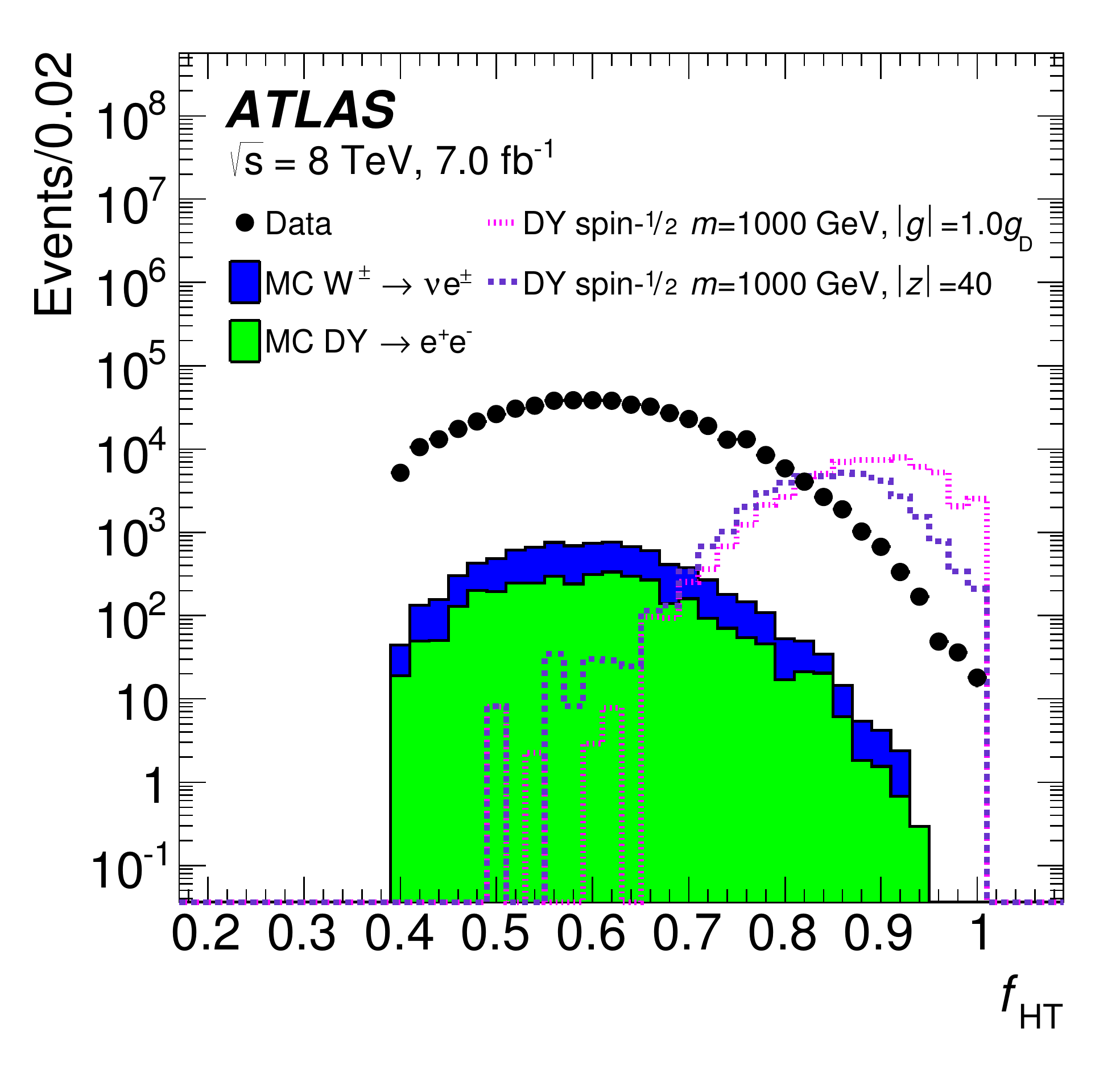}
\caption{Distributions of the EM energy dispersion $w$ (left) and fraction of TRT HT hits $f_{\rm HT}$ (right) at the last stage of the event selection (prior to the requirements on these two variables). Electroweak background MC samples with electrons in the final state (luminosity-weighted) as well as signal samples of various HIP charges and $m=1000$~GeV (luminosity-weighted $\times 500$) are also shown. Multijet processes (not simulated) are responsible for most of the candidates observed in data.} 
\label{fig:distributions}
\end{center}
\end{figure}

The last two selection criteria on $w$ and $f_{\rm HT}$ are very effective at reducing backgrounds and at the same time retaining potential signals, as shown in Table~\ref{tab:cutflow} and in Fig.~\ref{fig:distributions}. 
These two variables are only slightly correlated, such that control regions for the data-driven background estimate can be defined (see Section~\ref{sec:backgrounds} and Fig.~\ref{fig:ABCD}).   
The EM dispersion $w$ is independent of the HIP mass and charge due to
the absence of an EM shower.  The energy loss of a HIP is
proportional to the square of the charge. Thus, HIPs with higher
charge produce more TRT HT hits, yielding a higher $f_{ \rm HT}$. No
significant dependence on the HIP mass is expected for $f_{\rm HT}$.

\begin{table}
\begin{center}
\caption{Number of events at each stage of the selection in data and in representative simulated signal samples (DY spin-1/2, $m=1000$~GeV and charges $|g|=1.0g_{\rm D}$ and $|z|=40$). See text for descriptions of the selection criteria. The percentages given in parentheses are relative efficiencies with respect to previous lines. }
\label{tab:cutflow}
\begin{tabular}{|l|c|c|c|c|c|}
\hline
                        & Data  & $|g|=1.0g_{\rm D}$   & $|z|=40$ \\
\hline
Total MC                & ---              & 26\,502                  & 23\,848                 \\
Level-1 trigger         & ---              & 7962 (30.0\%)          & 6319 (26.5\%)        \\
HIP trigger             & 854\,130           & 6526 (82.0\%)          & 4481 (70.9\%)         \\
Preselection            & 600\,358 (70.3\%)  & 6503 (99.7\%)          & 4431 (98.9\%)         \\
EM layers               & 591\,627 (98.5\%)  & 6503 (100\%)           & 4421 (99.8\%)         \\
Pseudorapidity          & 501\,304 (84.7\%)  & 6242 (96.0\%)          & 4072 (92.1\%)         \\
Hadronic veto           & 498\,993 (99.5\%)  & 6242 (100\%)           & 4071 (100\%)         \\
EM dispersion           & 3                & 6224 (99.7\%)          & 4065 (99.9\%)         \\
TRT HT hits             & 0                & 6195 (99.5\%)          & 4018 (98.8\%)         \\
\hline
\end{tabular}
\end{center}
\end{table}

\subsection{Selection efficiencies in fiducial kinematic regions}
\label{sec:eff_fiducial}

Following the example of previously published ATLAS HIP searches~\cite{ATLAS2011a,ATLAS2012a}, fiducial regions of the HIP kinematic parameter space are identified in which the selection efficiency is high and uniform. This permits an interpretation of the results that does not depend on the assumed model of HIP production. The fiducial regions can be defined in terms of HIP kinetic energy and pseudorapidity and need to be determined separately for each value of HIP charge and mass, using the fully simulated single-particle HIP samples described in Section~\ref{sec:simulations}. Since the efficiency within the region is uniform by definition, the search results can then be interpreted in any model of HIP production by counting the number of events within the region.

The minimum particle kinetic energy to which the search is sensitive depends on the amount of material that a HIP needs to traverse before reaching the EM calorimeter. The maximum energy depends on the amount of material before the HIP reaches the hadronic calorimeter (where it provokes the hadronic veto of the Level-1 trigger). From simple geometric considerations, in the EM barrel, this material is roughly proportional to $(\sin\theta)^{-1}$, while in the EM endcap it varies as $(\cos\theta)^{-1}$. Therefore, the $\eta$ dependence of the minimum and maximum energy values can be canceled out to first order by defining them in terms of transverse kinetic energy \mbox{($E^{\rm kin}_{\rm T}=E_{\rm kin} \sin\theta$)} in the EM barrel region \mbox{($|\eta|<1.475$)} and longitudinal kinetic energy  \mbox{($E^{\rm kin}_{\rm L}=E_{\rm kin} \cos\theta$)} in the EM endcap region \mbox{($|\eta|>1.475$)}.

\begin{figure}[ht]
  \begin{center}
    \includegraphics[width=0.4\linewidth]{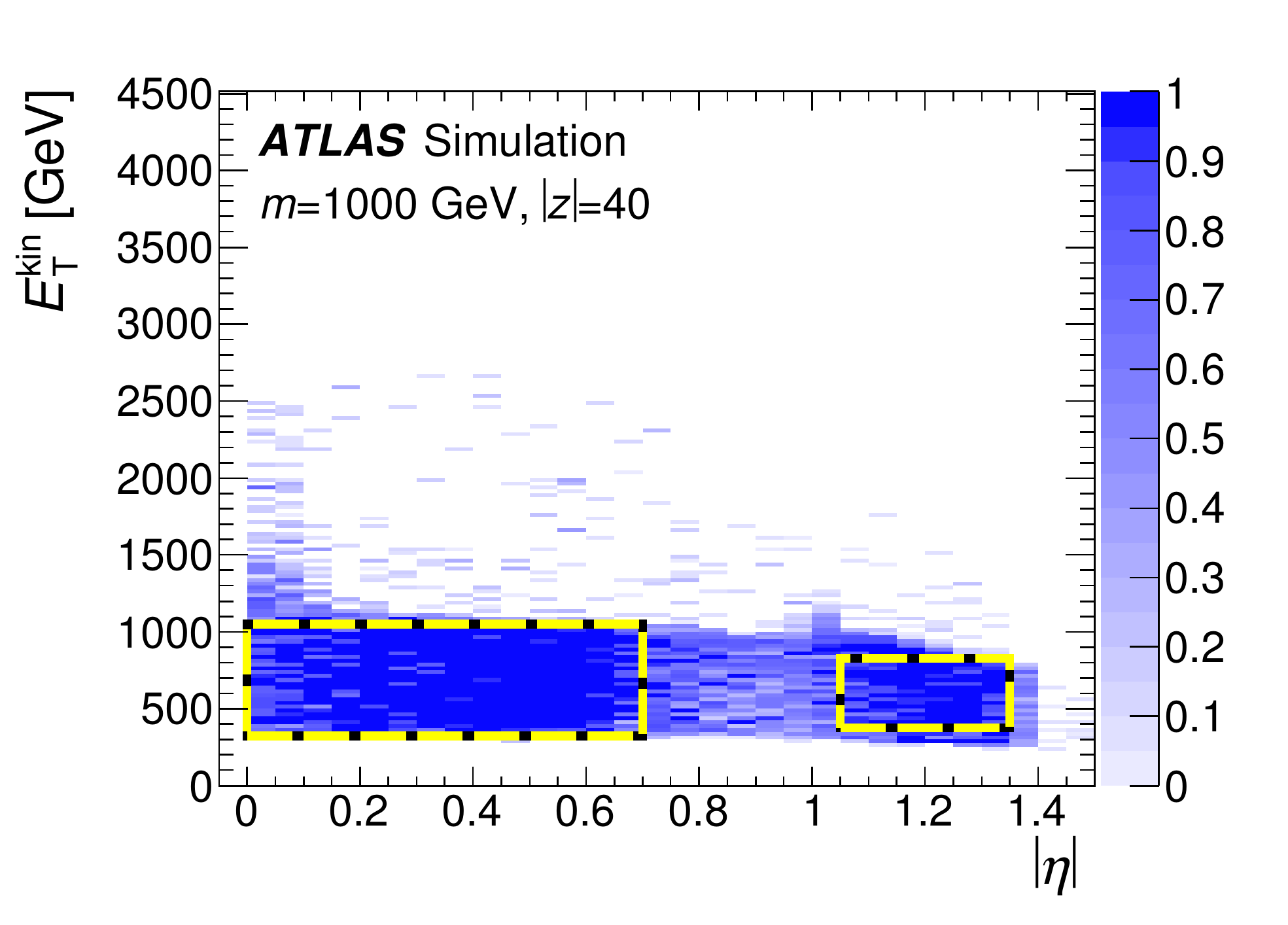}
    \includegraphics[width=0.4\linewidth]{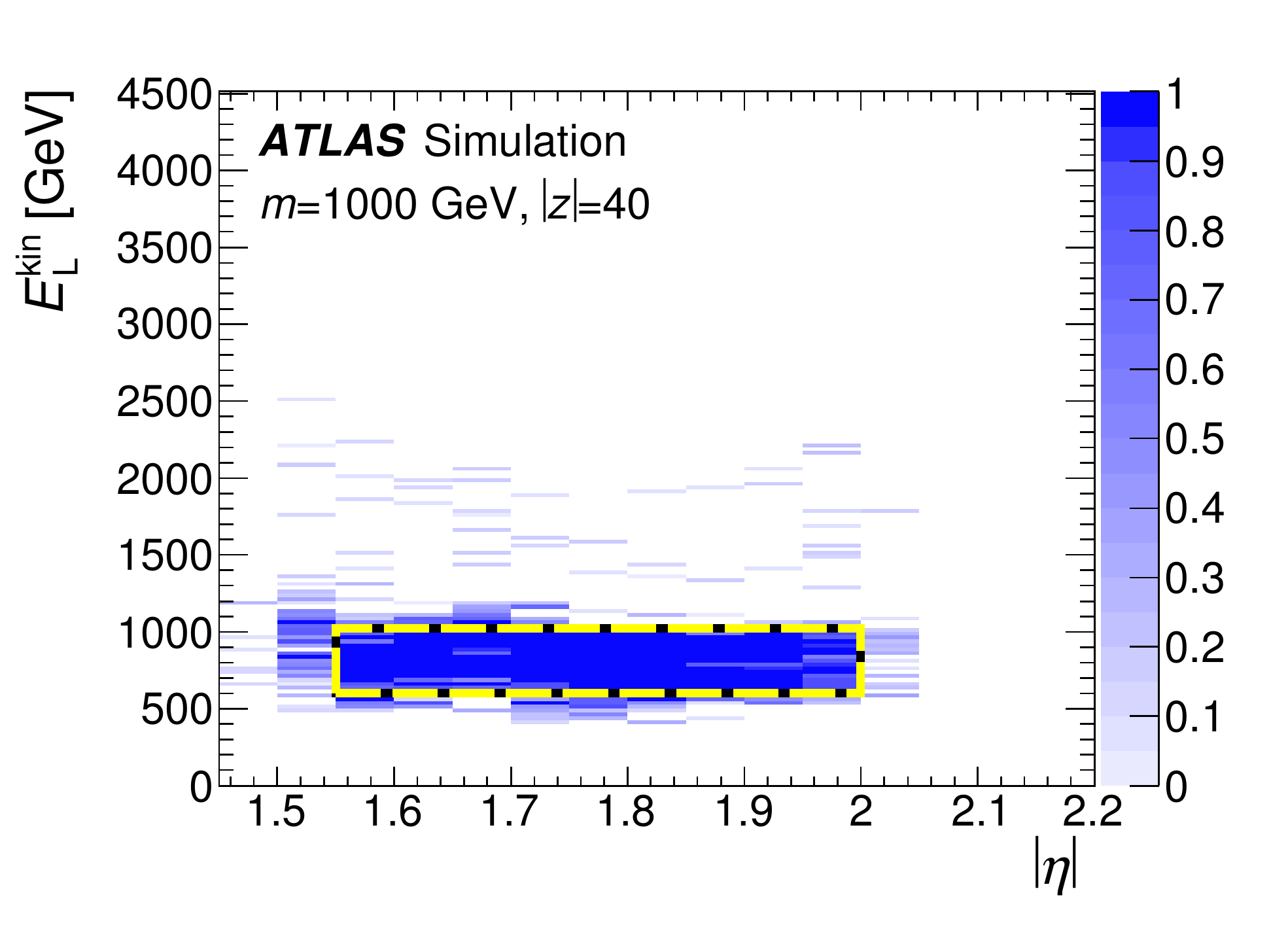}
    \includegraphics[width=0.4\linewidth]{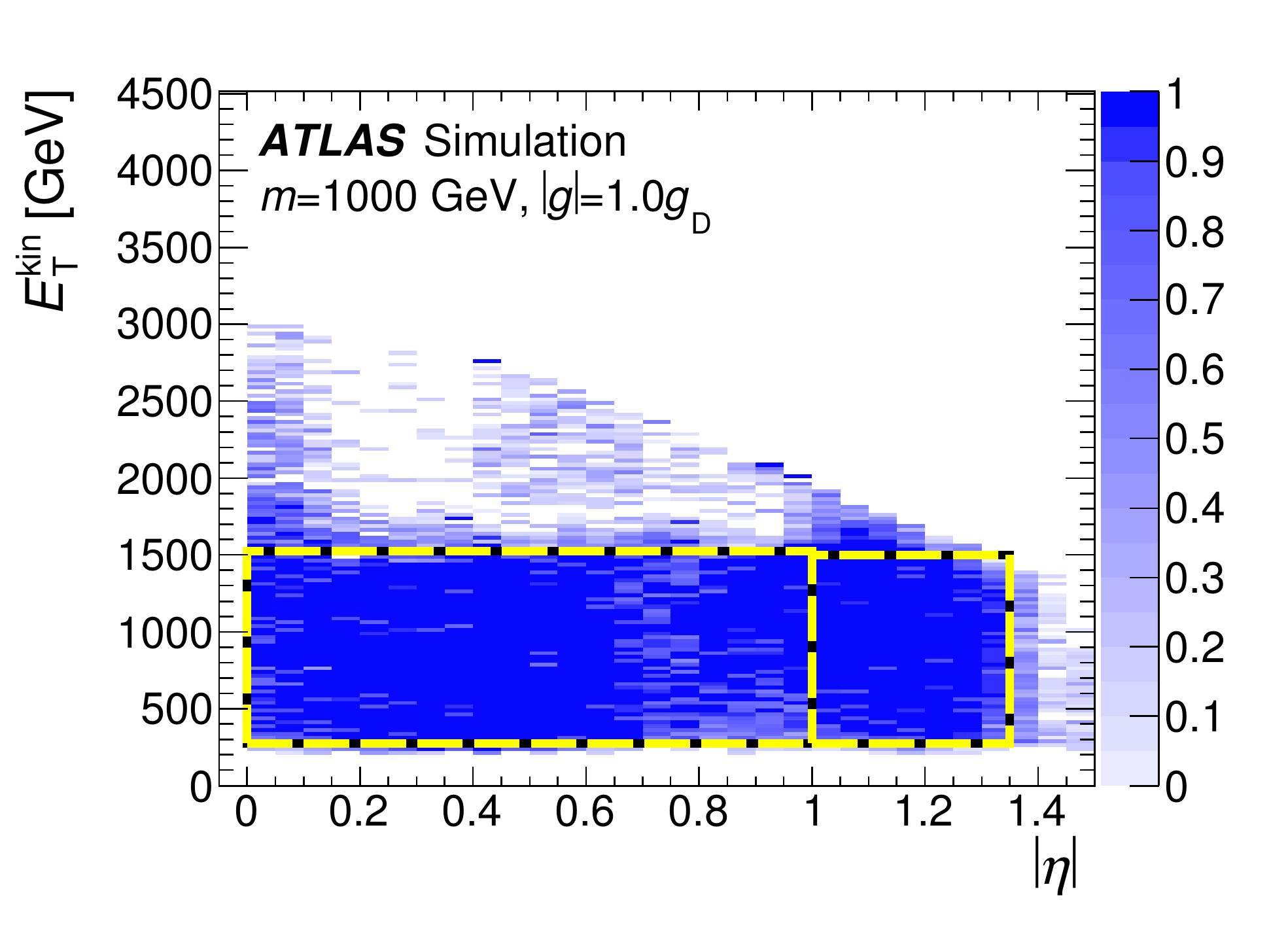}
    \includegraphics[width=0.4\linewidth]{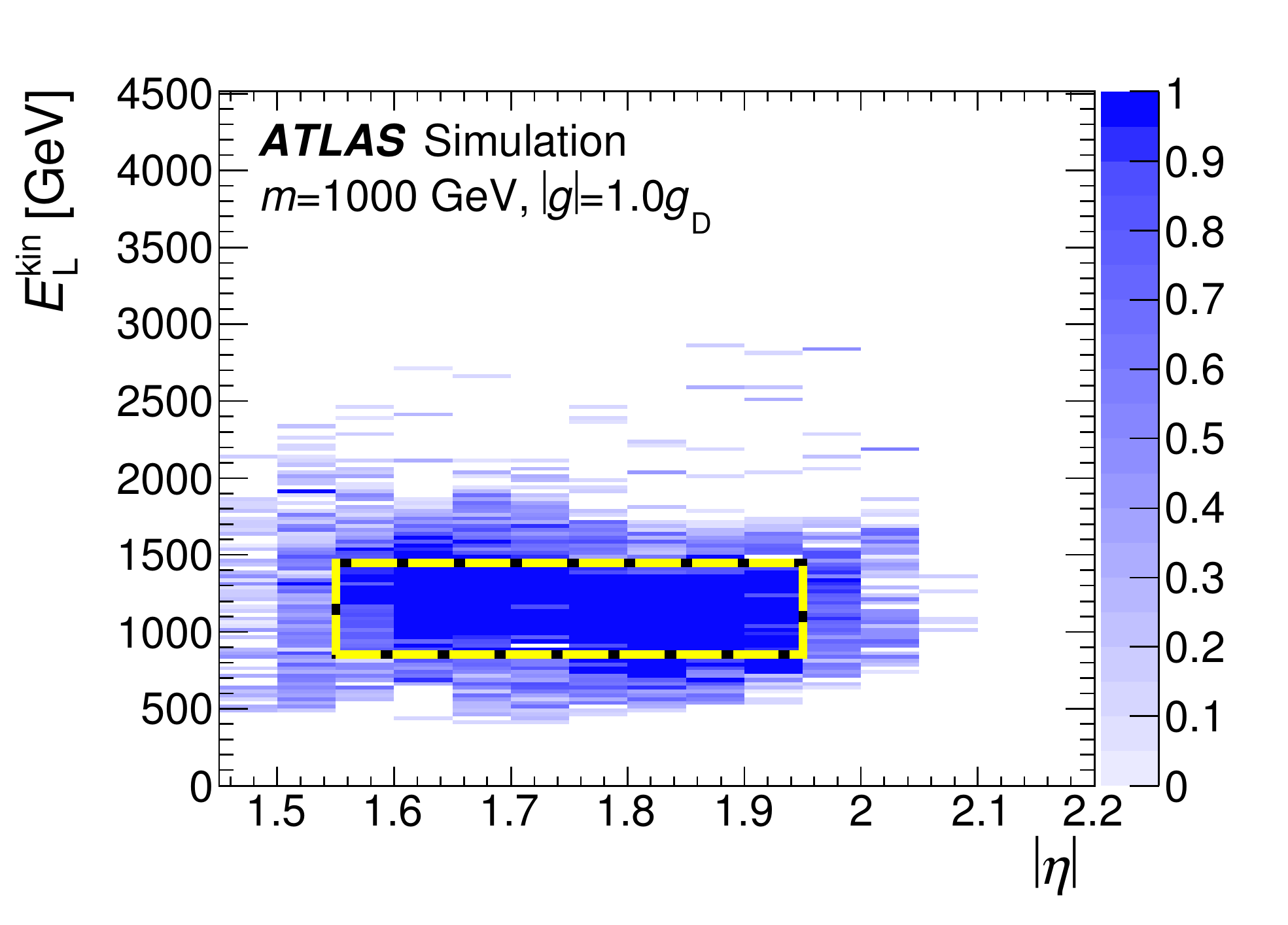}
    \includegraphics[width=0.4\linewidth]{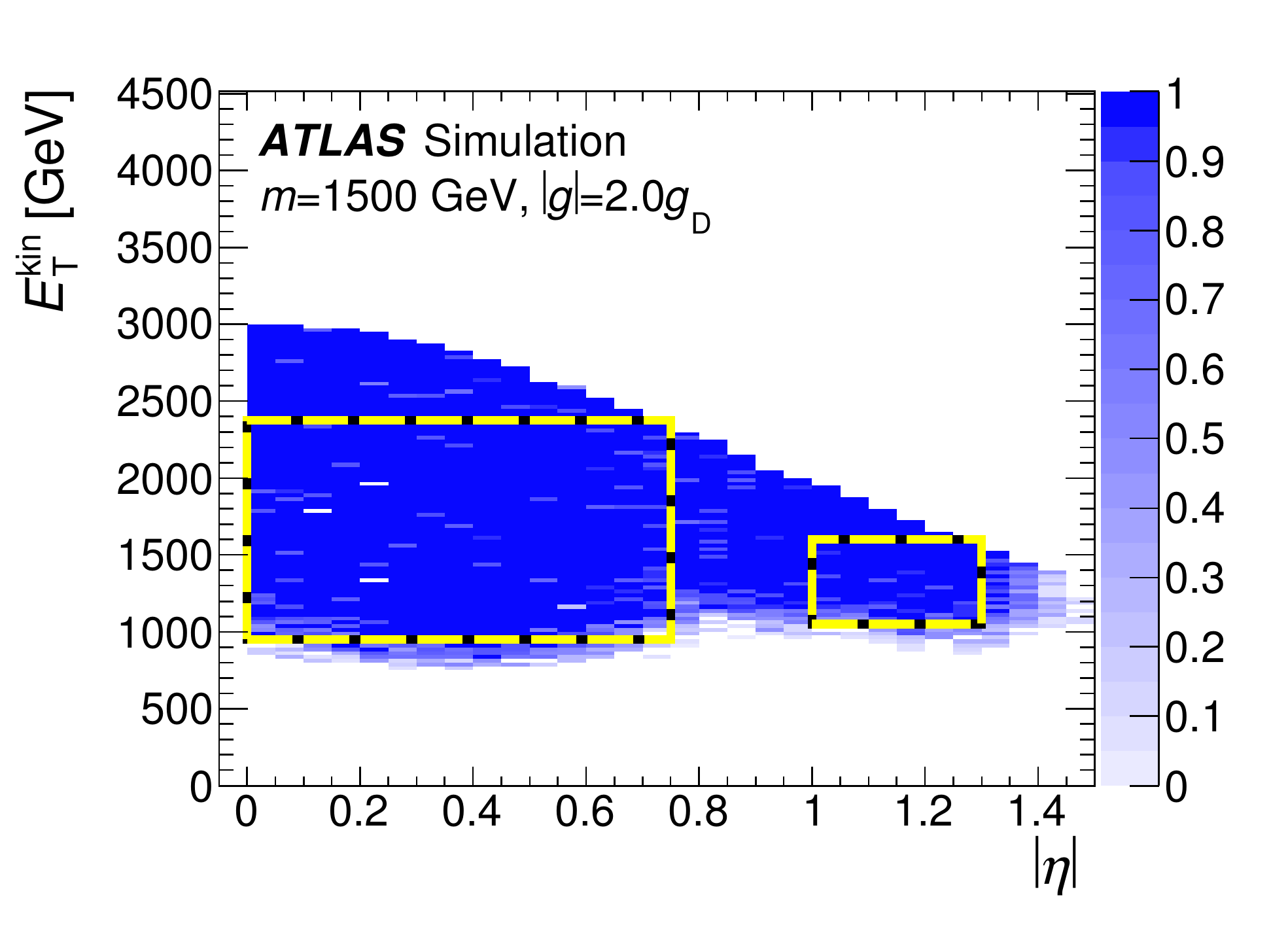}
    \includegraphics[width=0.4\linewidth]{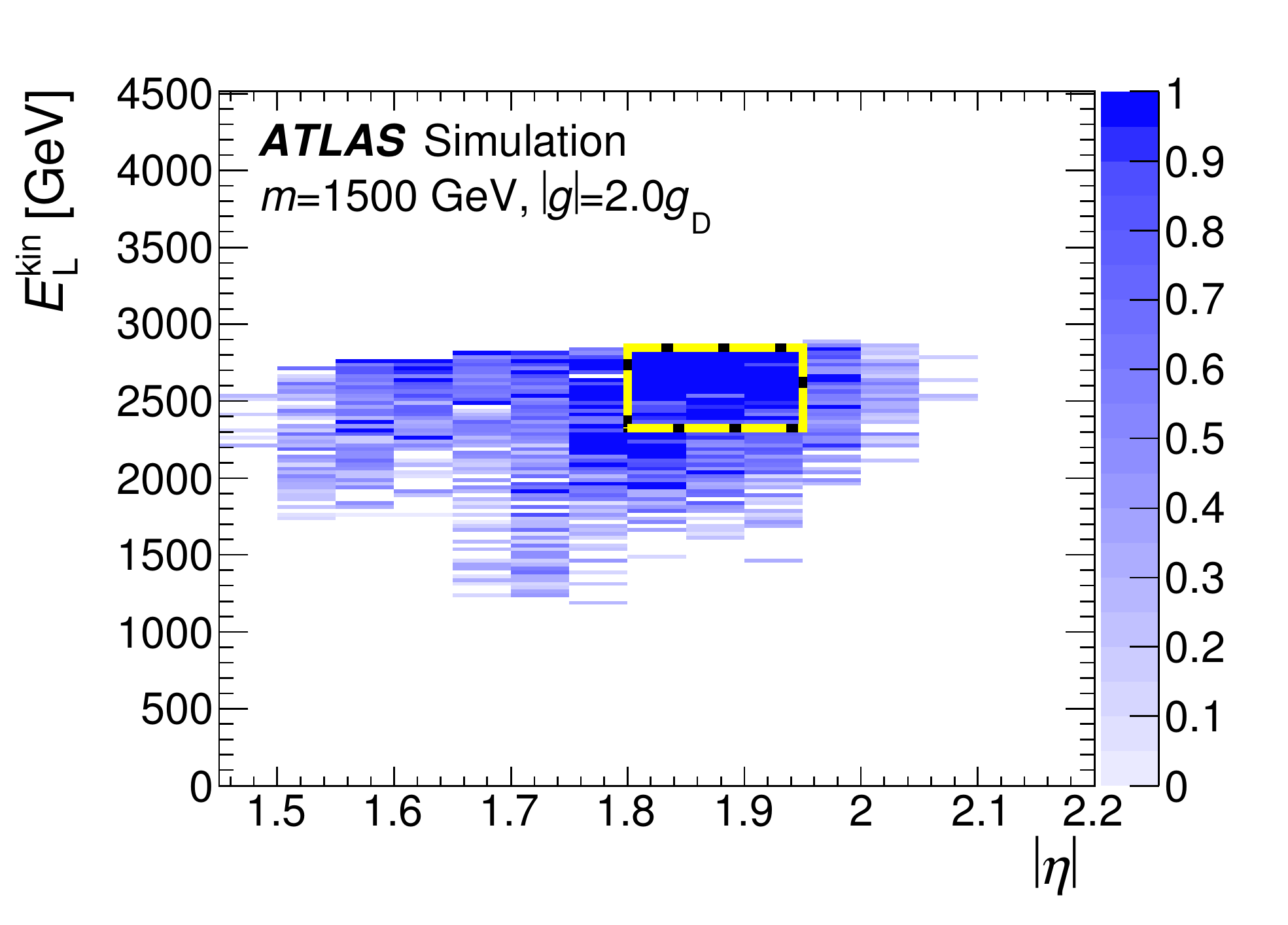}
  \end{center}
  \caption{Total selection efficiency (i.e., the fraction of MC events surviving all the criteria listed in Table~\ref{tab:cutflow}) as a function of transverse kinetic energy (left) or longitudinal kinetic energy (right) and pseudorapidity, for HIPs with mass 1000~GeV and charge $|z|=40$ (top), mass 1000~GeV and charge $|g|=1.0g_{\rm D}$ (middle) and mass 1500~GeV and charge $|g|=2.0g_{\rm D}$ (bottom). These plots are obtained using fully simulated single-particle samples with a uniform kinetic energy distribution between 0 and 3000~GeV. The fiducial regions (as defined in the text) are indicated by rectangular dashed lines.}
  \label{fig:acceptances}
\end{figure}

As can be seen in Fig.~\ref{fig:acceptances} in the case of three representative signals, fiducial regions in the $E^{\rm kin}_{\rm T}$ versus $|\eta|$ plane appear as rectangles for the EM barrel region. Likewise, rectangles can be defined in the $E^{\rm kin}_{\rm L}$ versus $|\eta|$ plane for the EM endcap regions. The reduced efficiency in the TRT barrel-endcap transition region \mbox{($0.77<|\eta|<1.06$)} visible in Fig.~\ref{fig:acceptances} (top left) motivates the consideration of a third region between $|\eta|=1.0$ and the end of the EM calorimeter barrel. 

\begin{figure}[ht]
  \begin{center}
    \includegraphics[width=0.49\linewidth]{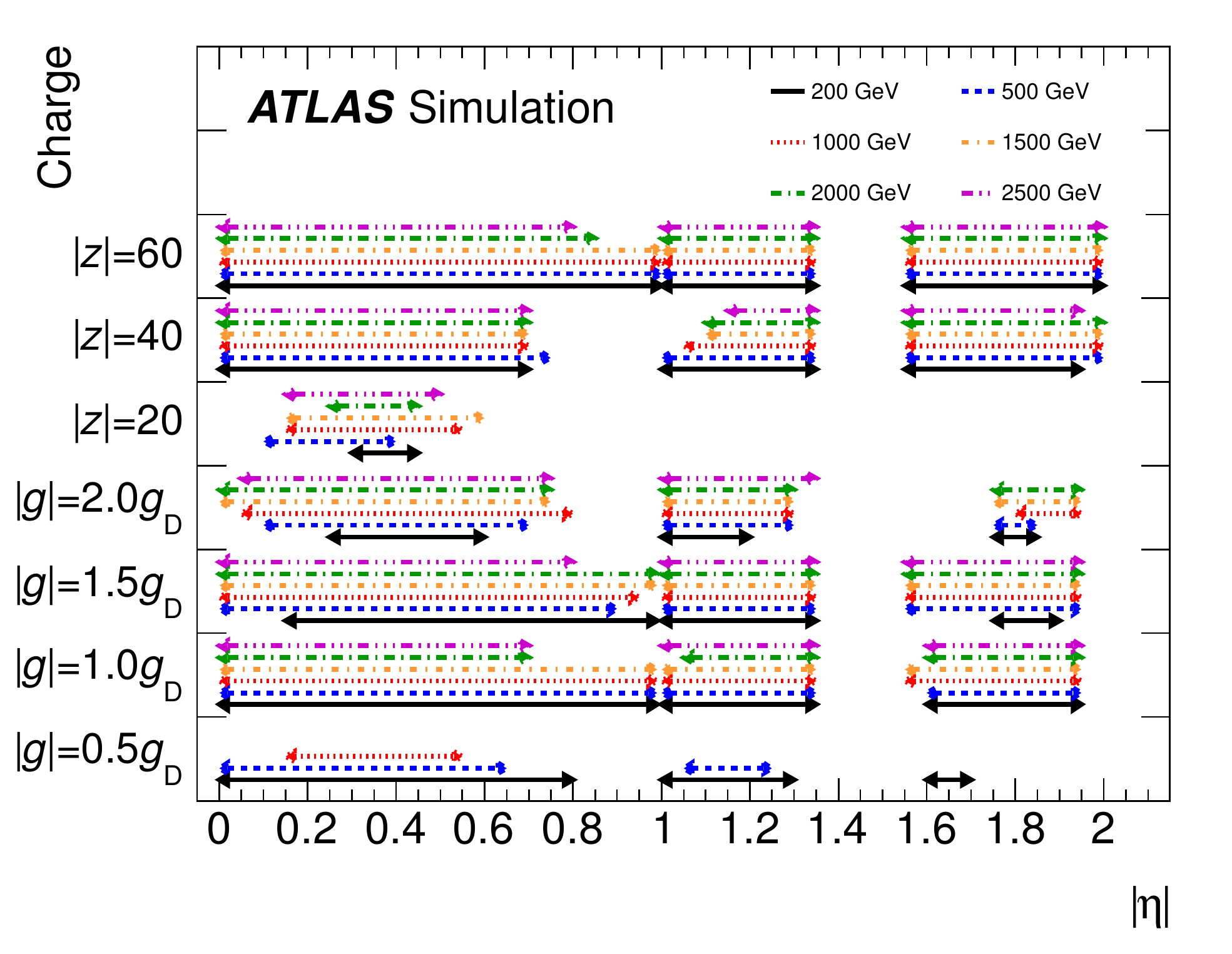}
    \includegraphics[width=0.49\linewidth]{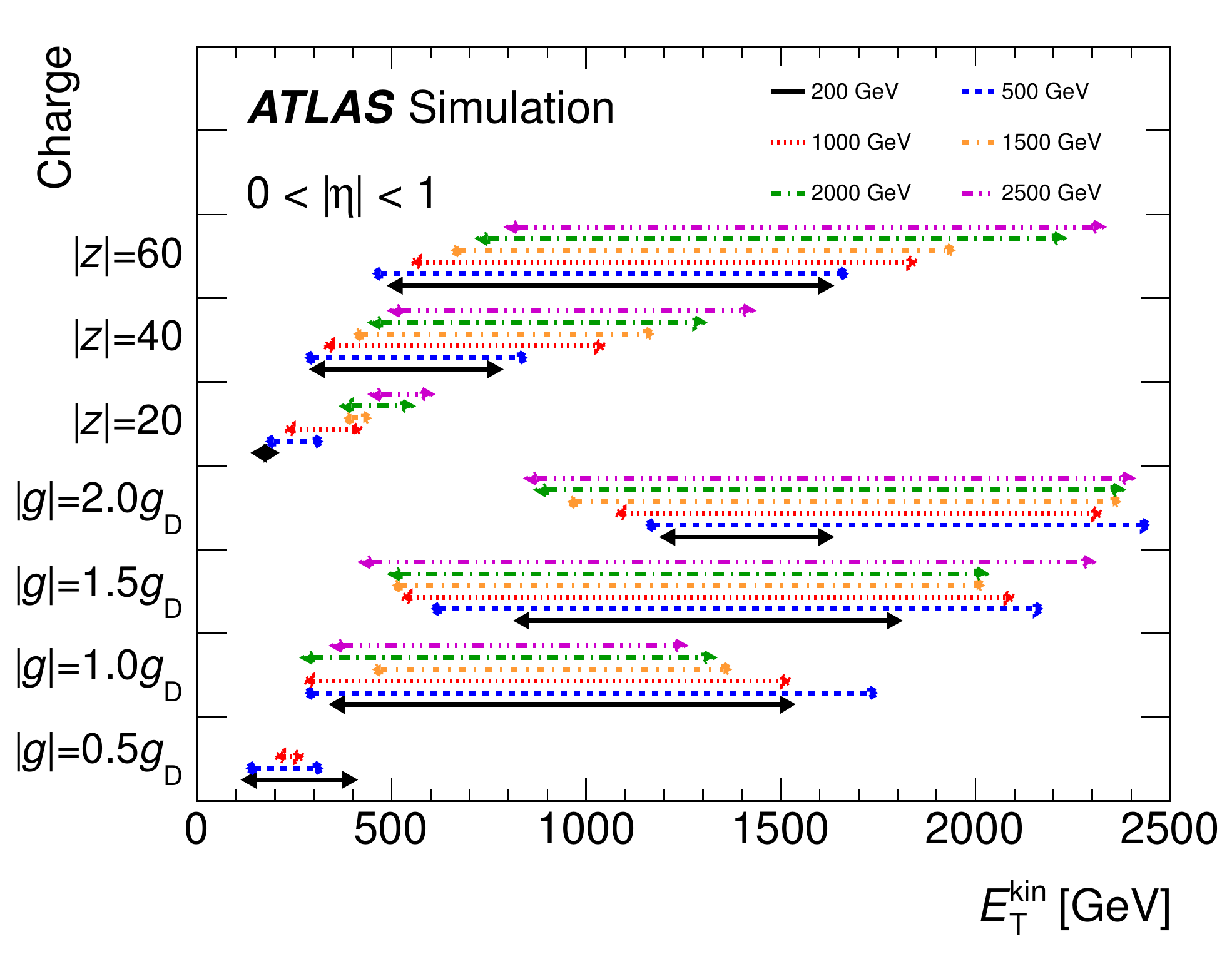}
    \includegraphics[width=0.49\linewidth]{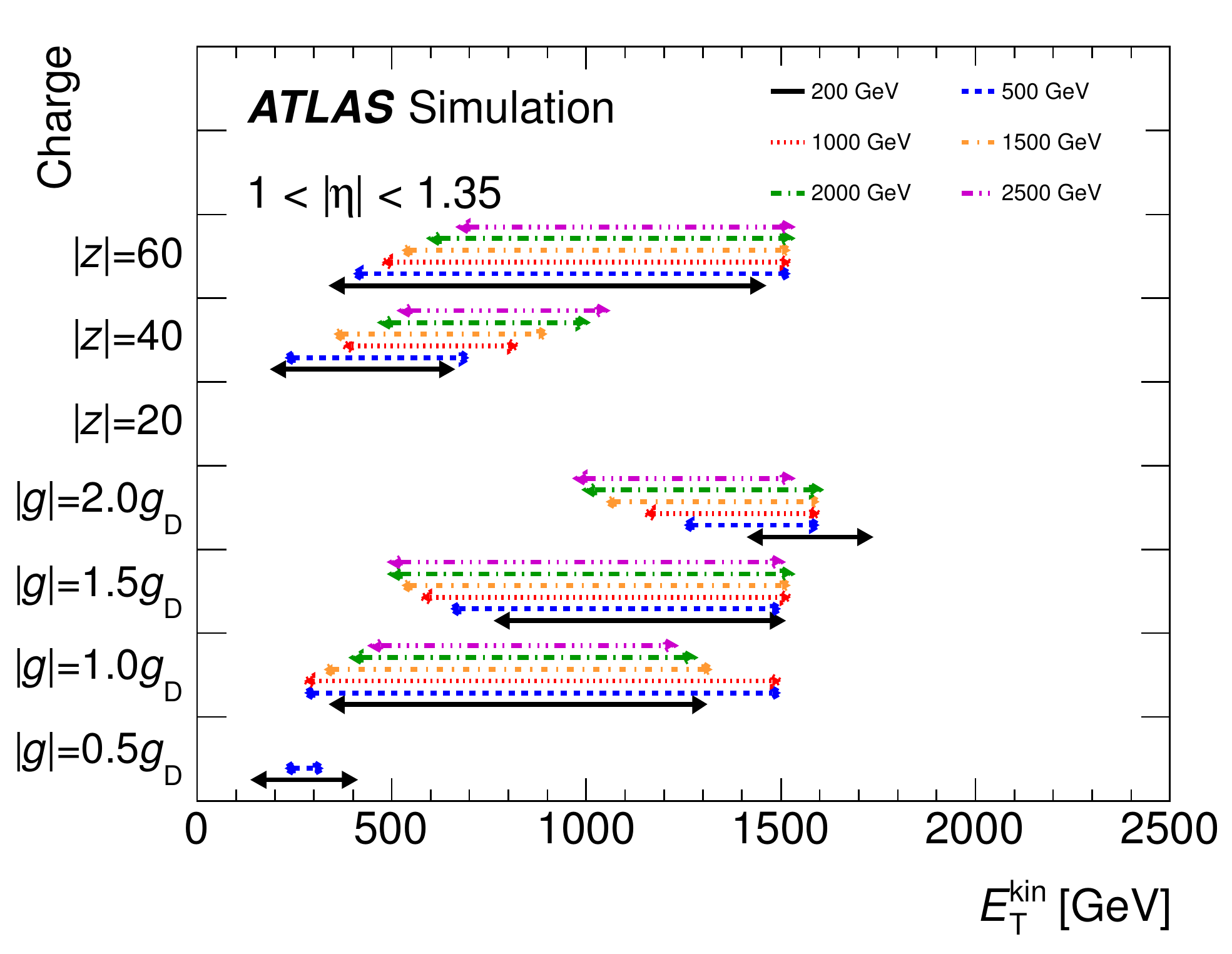}
    \includegraphics[width=0.49\linewidth]{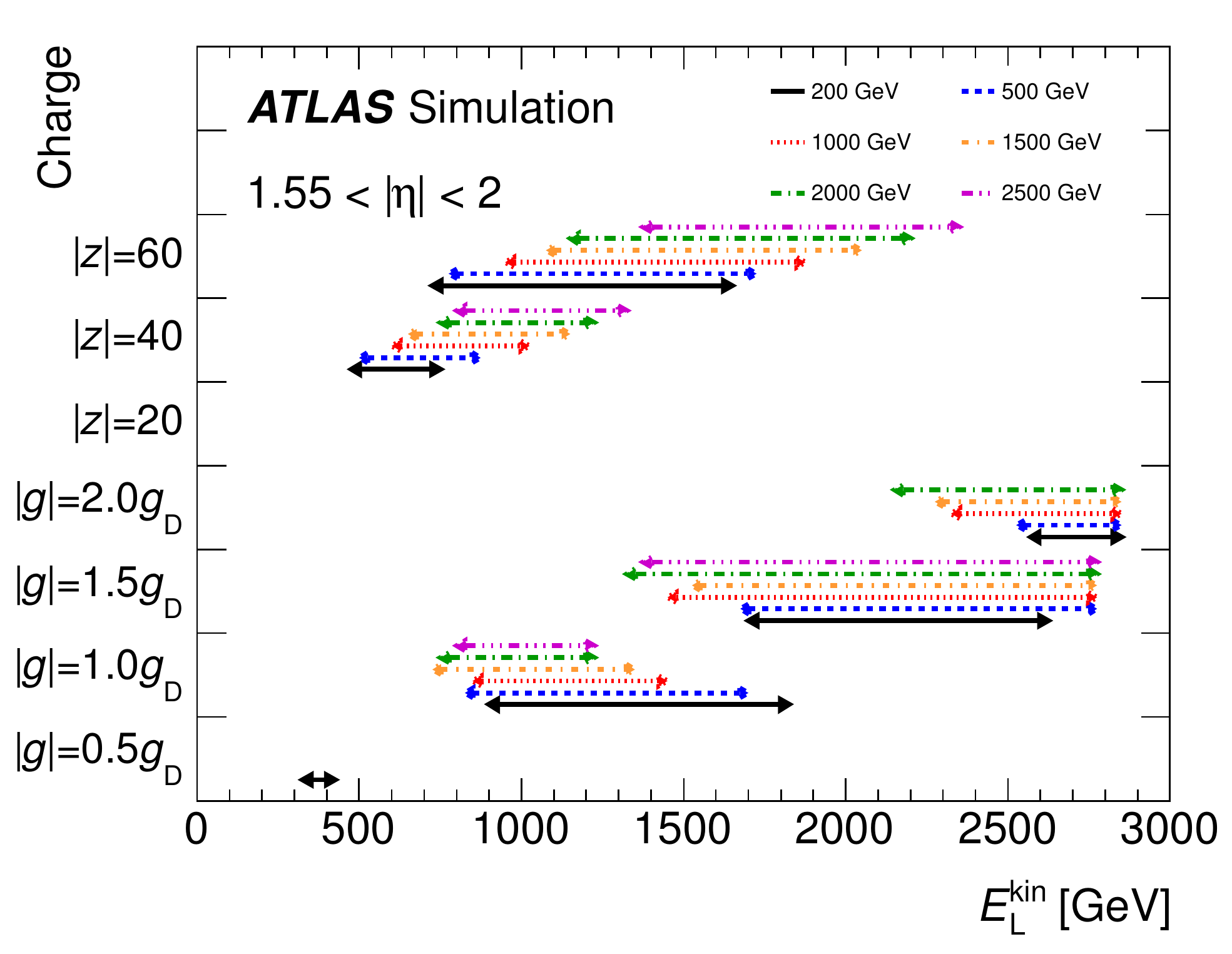}
  \end{center}
  \caption{Fiducial regions for the HIP charges considered in the search, as defined in Section~\ref{sec:eff_fiducial}. The various line styles correspond to different HIP masses. The top left plot shows the $|\eta|$ acceptance ranges, while the other plots show the $E^{\rm kin}$ acceptance ranges corresponding to the three different $|\eta|$ ranges. Blank space means that no fiducial region of high efficiency is found for the corresponding mass and charge.}
  \label{fig:fiducial}
\end{figure}

The rectangles that define the fiducial regions are determined by first dividing the $E^{\rm kin}_{\rm T}$ ($E^{\rm kin}_{\rm L}$ for the EM endcaps) versus $|\eta|$ plane into bins of size 25~GeV~$\times$~0.05 and using an algorithm that identifies the largest rectangular region for which the average selection efficiency across all bins inside the region is larger than 90\% with a standard deviation lower than 12.5\%. The value of the standard deviation cut was chosen as a compromise between performance of the algorithm and a well-defined efficiency of a region. For some mass and charge points, such regions are too narrow to be found with this definition, hence, no model-independent cross section limit is obtained for those points. In particular, no fiducial region was found for HIPs with electric charge $|z|=10$ for any mass point.
Figure~\ref{fig:fiducial} shows the various identified fiducial regions in $|\eta|$ (top left) as well as the regions in $E_{\rm T}^{\rm kin}$ corresponding to the two $|\eta|$ regions in the barrel (top right and bottom left) and the regions in $E_{\rm L}^{\rm kin}$ corresponding to the $|\eta|$ region in the endcap (bottom right), for all relevant mass and charge points.

\subsection{Selection efficiencies in pair-production models}
\label{sec:eff_pair}

Fully simulated events are used to determine selection efficiencies for a DY fermion (spin-1/2) pair-production process for electric as well as magnetic charges.
The selection efficiencies for spin-0 DY HIPs, which were not fully simulated, are determined as follows: fine efficiency maps (finely binned in kinetic energy and pseudorapidity) were obtained from fully simulated single-particle samples and folded with the generator-level spin-0 DY angular distributions. As a cross-check, the same method applied to spin-1/2 DY HIPs was found to give results no more than 9\% discrepant from those obtained using the fully simulated spin-1/2 DY sample.

As discussed in Section~\ref{sec:trigger_performance}, the main losses in all cases are due to the acceptance of the Level-1 trigger. In particular, for high charges, a large fraction of the HIPs produced in DY events lose all their energy and stop before they reach the EM calorimeter. The
acceptance for DY-produced monopoles with charge $|g|=2.0g_{\rm D}$ is very small, of the order of 0.1\%. For this charge, the ionization energy loss is such that only monopoles with transverse energy higher than $\sim$1200~GeV in the barrel and longitudinal energy higher than $\sim$1500~GeV in the endcap have a chance to pass the Level-1 trigger. Such energies lie in the extreme tails of the 8~TeV DY pair-production energy distributions. High-charge HIPs thus have low acceptances, which are highly dependent on the tails of the distributions, and hence very model-dependent. For this reason, the search is not interpreted for DY signals with acceptances lower than 1\%. This includes all $|g|=2.0g_{\rm D}$ mass points as well as the $|g|=1.5g_{\rm D}$, $m=200$~GeV point and the $|z|=60$, $m=2000$~GeV and $m=2500$~GeV points. 

Full selection efficiencies are presented in Table~\ref{tab:eff_models} for spin-1/2 and spin-0 HIPs in the DY production model for all masses and charges considered in the search. 
The mass dependence comes from differences in energy and angular distributions, and also from the velocity dependence of the energy loss, as more massive HIPs have lower $\beta$ on average, which leads to lower energy loss for monopoles (or generally higher energy loss for electrically charged particles). 
Spin-0 HIPs have a higher acceptance due to the narrower angular distribution~\cite{MadGraph}.

\begin{table}
\caption{Event selection efficiencies (i.e., the fraction of MC events surviving all the criteria listed in Table~\ref{tab:cutflow}) in \% for spin-1/2 (top) and spin-0 (bottom) HIPs with DY production kinematic distributions. The quoted uncertainties are due to MC sample size.}
\label{tab:eff_models}
\small
\begin{center}
\begin{tabular}{|r|c|c|c|c|c|c|c|}
    \hline
    $m$ [GeV]                      & $|g|=0.5g_{\rm D}$ & $|g|=1.0g_{\rm D}$    & $|g|=1.5g_{\rm D}$    & $|z|=10$     & $|z|=20$     & $|z|=40$     & $|z|=60$      \\
    \hline
    spin-1/2 &&&&&&& \\
    200 & 22.3 $\pm$ 0.3 & 3.5 $\pm$ 0.1 & 0.14 $\pm$ 0.03 & 3.8 $\pm$ 0.1 & 9.7 $\pm$ 0.2 & 11.9 $\pm$ 0.2 & 3.1 $\pm$ 0.1 \\   
    500 & 33.5 $\pm$ 0.3 & 14.9 $\pm$ 0.3 & 1.16 $\pm$ 0.09 & 6.7 $\pm$ 0.2 & 19.0 $\pm$ 0.3 & 20.0 $\pm$ 0.3 & 6.2 $\pm$ 0.2 \\
   1000 & 27.8 $\pm$ 0.3 & 23.4 $\pm$ 0.3 & 3.7 $\pm$ 0.1 & 10.7 $\pm$ 0.2 & 24.6 $\pm$ 0.3 & 16.9 $\pm$ 0.3 & 3.8 $\pm$ 0.1 \\ 
   1500 & 23.7 $\pm$ 0.3 & 22.2 $\pm$ 0.3 & 3.5 $\pm$ 0.1 & 13.8 $\pm$ 0.2 & 22.5 $\pm$ 0.3 & 10.0 $\pm$ 0.2 & 1.43 $\pm$ 0.09 \\
   2000 & 16.7 $\pm$ 0.3 & 16.5 $\pm$ 0.3 & 2.8 $\pm$ 0.1 & 15.5 $\pm$ 0.3 & 17.5 $\pm$ 0.3 & 3.7 $\pm$ 0.1 & 0.24 $\pm$ 0.03 \\  
   2500 & 9.8 $\pm$ 0.2 & 9.8 $\pm$ 0.2 & 1.61 $\pm$ 0.09 & 12.3 $\pm$ 0.2 & 10.2 $\pm$ 0.2 & 1.05 $\pm$ 0.07 & 0.009 $\pm$ 0.007 \\
    \hline
    spin-0 &&&&&&& \\
    200 & $42.5 \pm 0.3$ & $10.0 \pm 0.2$ & $0.40\pm 0.04$ & $5.9 \pm 0.2$ & $28.0 \pm 0.3$ & $27.6\pm 0.3$ & $8.2 \pm 0.2$ \\
    500 & $53.8 \pm 0.3$ & $34.8 \pm 0.3$ & $4.1 \pm 0.1$  & $9.8 \pm 0.2$ & $35.3 \pm 0.3$ & $42.1 \pm 0.3$ & $15.1 \pm 0.2$ \\
   1000 & $44.3 \pm 0.3$ & $51.1 \pm 0.3$ & $11.4 \pm 0.2$ & $15.1 \pm 0.2$ & $45.7 \pm 0.3$ & $37.5 \pm 0.3$ & $11.4 \pm 0.2$ \\
   1500 & $36.5 \pm 0.3$ & $49.7 \pm 0.3$ & $13.8 \pm 0.2$ & $19.9 \pm 0.3$ & $47.7 \pm 0.3$ & $26.7 \pm 0.3$ & $4.8 \pm 0.1$ \\
   2000 & $30.9 \pm 0.3$ & $41.6 \pm 0.3$ & $10.9 \pm 0.2$ & $25.5 \pm 0.3$ & $43.6 \pm 0.3$ & $13.2 \pm 0.2$ & $1.15 \pm 0.07$\\
   2500 & $22.9 \pm 0.3$ & $30.8 \pm 0.3$ & $6.9 \pm 0.2$  & $26.9 \pm 0.3$ & $31.7 \pm 0.3$ & $4.3 \pm 0.1$  & $0.18 \pm 0.03$\\
    \hline
\end{tabular}
\end{center}
\normalsize
\end{table}

\section{Background estimate}
\label{sec:backgrounds}

The selection criteria defined in Section~\ref{sec:selection} efficiently reject Standard Model backgrounds. In particular, the vast majority of EM cluster candidates in multijet events feature broad energy depositions in all three EM layers and few associated TRT HT hits. Jet backgrounds could pass the full selection in cases of extremely rare events in which the EM calorimeter shower shape is misreconstructed such as to appear very narrow in all EM layers and the trajectories of several charged particles overlap in the TRT to cross the same set of straws and produce HT hits. 
Processes featuring isolated electrons with transverse momenta exceeding the Level-1 trigger threshold can also constitute backgrounds, despite their lower cross sections. Those are largely dominated by $W$ and $Z$ production (described in Section~\ref{sec:simulations}). Electron showers are narrower than jets, and such processes lead to a reconstructed $w$ distribution that lies closer to the signal region, as can be seen in Fig.~\ref{fig:distributions}. Near the signal region, candidates from electrons from $W$ and $Z$ decays are comparable in yield to candidates from multijet events. Hot cells in the EM calorimeter do not constitute backgrounds as they are never found to be associated with TRT HT hits while remaining isolated. 

\begin{figure}[ht]
  \begin{center}
\includegraphics[width=0.7\linewidth]{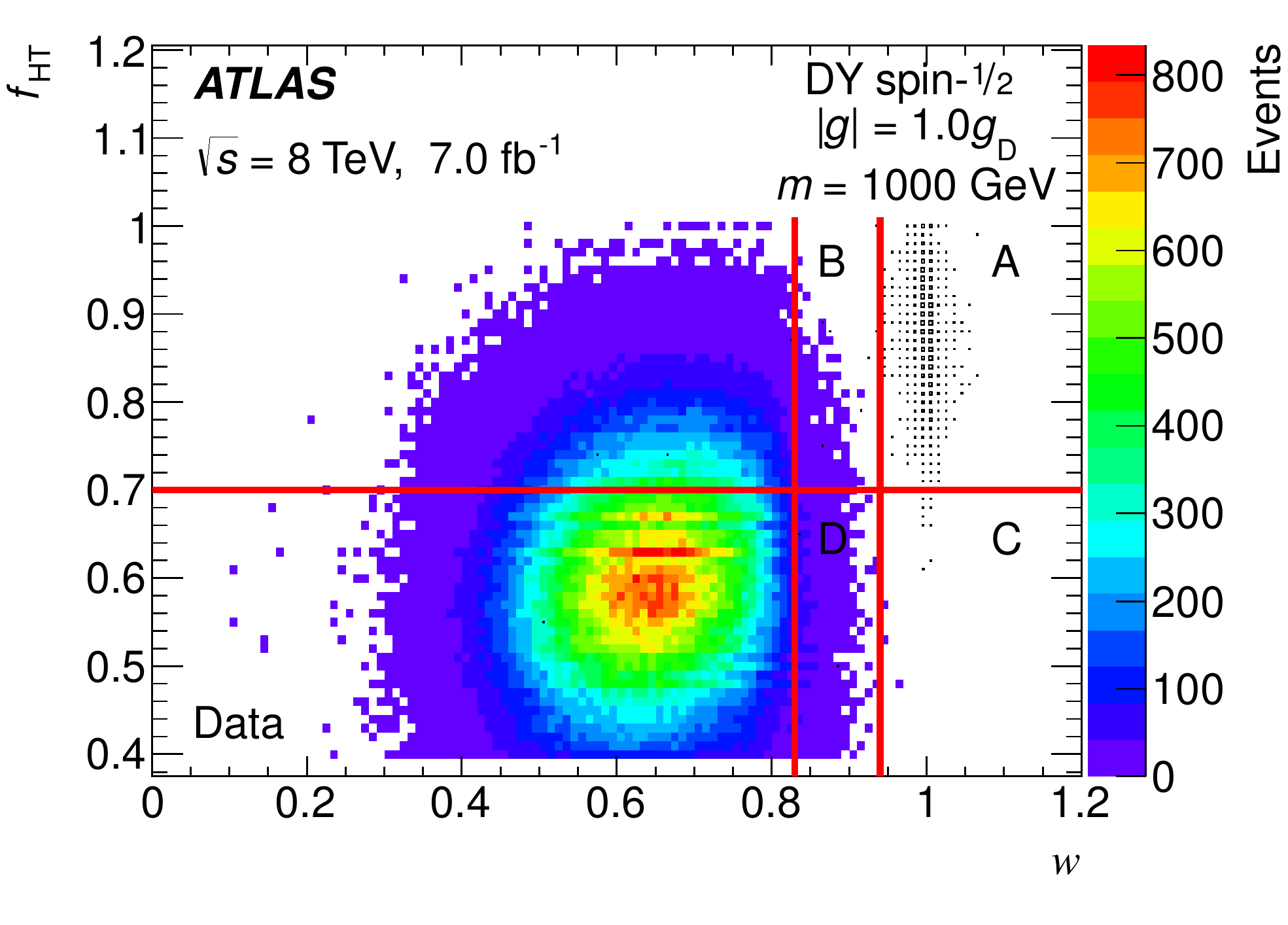}
  \end{center}
  \caption{Candidates seen in data (color scale) and in a representative simulated signal sample (black squares) in the $f_{\rm HT}$ versus $w$ plane, at the last stage of the event selection (prior to the requirements on these two variables). The number of background events in the signal region (A) is estimated using the left and bottom bands (B, D, and C) as control regions, as described in the text.}
  \label{fig:ABCD}
\end{figure}

A fully data-driven background estimate is performed in this search. 
This approach is necessary because it is unrealistic to produce the
enormous number of MC events required to model the QCD background, but
it also ensures that all possible background sources, including those
not foreseen, are taken into account.
The candidates passing the selection requirements except for the final EM dispersion and TRT HT hit criteria are shown in Fig.~\ref{fig:ABCD} in the plane defined by the two remaining discriminating variables, $f_{\rm HT}$ and $w$. This plane is divided into A, B, C, and D regions, where A is the signal region. The main assumption on which the background estimation method relies is that the ratio of region A to region C background events is the same as the ratio of region B to region D background events, or, in other words, that $f_{\rm HT}$ and $w$ are independent variables. Detector geometry effects give rise to a correlation due to the slight pseudorapidity ($|\eta|$) dependence of the $f_{\rm HT}$ and $w$ variables. The correlation is small near the signal region but increases somewhat at lower $w$ values. This motivates the choice of $w=0.84$ as the lower $w$ limit of the B and D control regions.  The lower $f_{\rm HT}$ limit of the C and D control regions is governed by the $f_{\rm HT}^{\rm trig}$ requirement applied by the Level-2 HIP trigger.  The absolute value of the Pearson correlation coefficient is below 0.05 in the control regions.
Given that the expected background is low, the correlations near the signal region are small, and the limited number of events precludes dividing the signal region into several separate $|\eta|$ regions, the data in the whole $|\eta|$ range are used without correction to estimate the backgrounds. The maximum possible difference between the ratios A/C and B/D due to correlations is estimated as follows. The B and D regions are extended to cover the range $0.69<w<0.91$ and divided into 22 narrow $w$ bins, with bin width chosen so as to provide sufficient statistics in each bin. 
The ratio B$_i$/D$_i$ is computed in each bin $i$. Taking as the weight the reciprocal square of the statistical uncertainty in each bin $j$ (such that $j>i$), the weighted average of the ratios B$_j$/D$_j$ across all bins $j>i$ is computed. This weighted average deviates from B$_i$/D$_i$ by no more than 40\%, which is taken as the systematic uncertainty in the background estimate obtained when assuming no correlations. 

Another concern is the possibility of signal contamination in the control regions. Contamination in B is negligible compared to background yields for all signal samples. However, contamination in C represents a significant fraction (more than $\sim 20\%$) of the signal for HIPs with low charges ($|g|=0.5g_{\rm D}$, $|z|=10$, and $|z|=20$), which produce fewer HT hits in the TRT on average. Before even knowing how many data events are observed in the signal region A, it is possible to estimate the expected limit on this number from a background estimate that takes signal contamination into account in a likelihood fit. Applying this method, it is found that signal contamination does not affect the expected limits in any significant way. As a cross-check, the expected number of background events in C is estimated by performing fits to the $w$ distribution observed in D assuming power-law and exponential functions, which both describe well the falling part of the distribution. Taking into account uncertainties obtained by using different functions and varying the fit parameters, the extrapolation predicts $4.7\pm1.0$ events in C, compatible with the three observed events. The fact that the three events in C do not appear at $w$ values near the peak of the signal distributions (at $w\sim1$) further supports the claim that they are not due to low-charge HIPs. 

The observed event yields in quadrants B, C, and D, are 626, 3, and 4615, respectively. 
The estimated number of background events in the signal region A, taking into account statistical uncertainties and systematic uncertainties due to possible correlations, is: 

$$ A_{\rm bkg}^{\rm est} = \frac{BC}{D} = 0.41 \pm 0.24 \rm{~(stat.)} \pm 0.16 \rm{~(syst.)}. $$

\section{Systematic uncertainties}
\label{sec:systematics}

Systematic uncertainties that can affect the estimated signal efficiencies are summarized below. These mostly concern possible imperfections in the description of the detector response to HIPs by the simulation.

\begin{itemize}

\item Electron-ion recombination effects in the sampling region of the EM calorimeter result in the loss of part of the energy deposition at high 
$dE/dx$ values. The fraction of visible energy is modeled in the ATLAS simulation using a modified Birks' law parameterization fitted to heavy-ion 
measurements in liquid argon~\cite{Burdin2012}. Varying the fraction of visible energy within its uncertainties results in a $\sim$10\% effect 
in efficiency for typical signals.

\item The fraction of HIPs that stop in the detector prior to reaching the EM calorimeter is affected by the assumed amount of material in the 
geometry description used by the {\sc GEANT4} simulation. Varying the simulated material density in the inner detector within the assumed uncertainties 
(which can range from $\pm 5\%$ to $\pm 15\%$~\cite{ATLAS2012e}) leads to a $\sim$5\% uncertainty in signal acceptance. This uncertainty is higher for 
charges $|g|=1.5g_{\rm D}$ and $|z|=10$ with a value of $\sim$10\%.

\item Secondary ionization by $\delta$-rays affects the TRT hit patterns. The kinetic energy threshold below which $\delta$-rays are not propagated explicitly in the ATLAS simulation
depends on the {\sc GEANT4} ``range cut'' parameter. Varying this parameter results in a $\sim$1\% uncertainty in the signal efficiency.

\item Pileup affects the efficiency as it adds a non-negligible number of TRT low-threshold hits inside the geometrical region considered for the $f_{\rm HT}^{\rm trig}$ and $f_{\rm HT}$ variables computed by the HIP trigger and the offline event selection, respectively. Uncertainties in pileup modeling and TRT hit occupancy result in $\sim$3\% uncertainty in the signal efficiency.

\item Cross-talk effects between EM calorimeter cells affect the $w_i$ variables and this is not fully described in the simulation. The resulting uncertainty in signal efficiency is $\sim$1\%.

\item For clusters delayed with respect to the expected arrival time of a highly relativistic particle by more than 10~ns, which corresponds to $\beta<0.37$, there is a significant chance that the event is triggered in the next bunch crossing by the Level-1 EM trigger. However, since HIP candidates selected by the L1 trigger necessarily have $\beta>0.4$, there are no significant losses (and no systematic uncertainties) due to timing effects.

\item An uncertainty in the efficiency of $\sim$1--3\% accounts for the statistical uncertainty from the MC signal samples.

\item For spin-0 DY HIPs, the relative uncertainty in efficiency due to the fact that efficiency maps are used instead of a full simulation is $\sim$9\%.

\end{itemize}

In addition to the uncertainties listed above, the systematic uncertainty due to the luminosity measurement is 2.8\%. It is derived following the same methodology as that detailed in Ref.~\cite{ATLASLUMI2013}. 

\section{Results}
\label{sec:result}

Zero events are observed in the signal region in 7.0~fb$^{-1}$ of 8~TeV proton-proton collision data, consistent with the background estimate. This results in an upper limit on the number of signal events of 3.0 at 95\% confidence level in the data sample.

\subsection{Cross section limits}
\label{sec:xs}

\begin{figure}[ht]
  \begin{center}
    \includegraphics[width=0.49\linewidth]{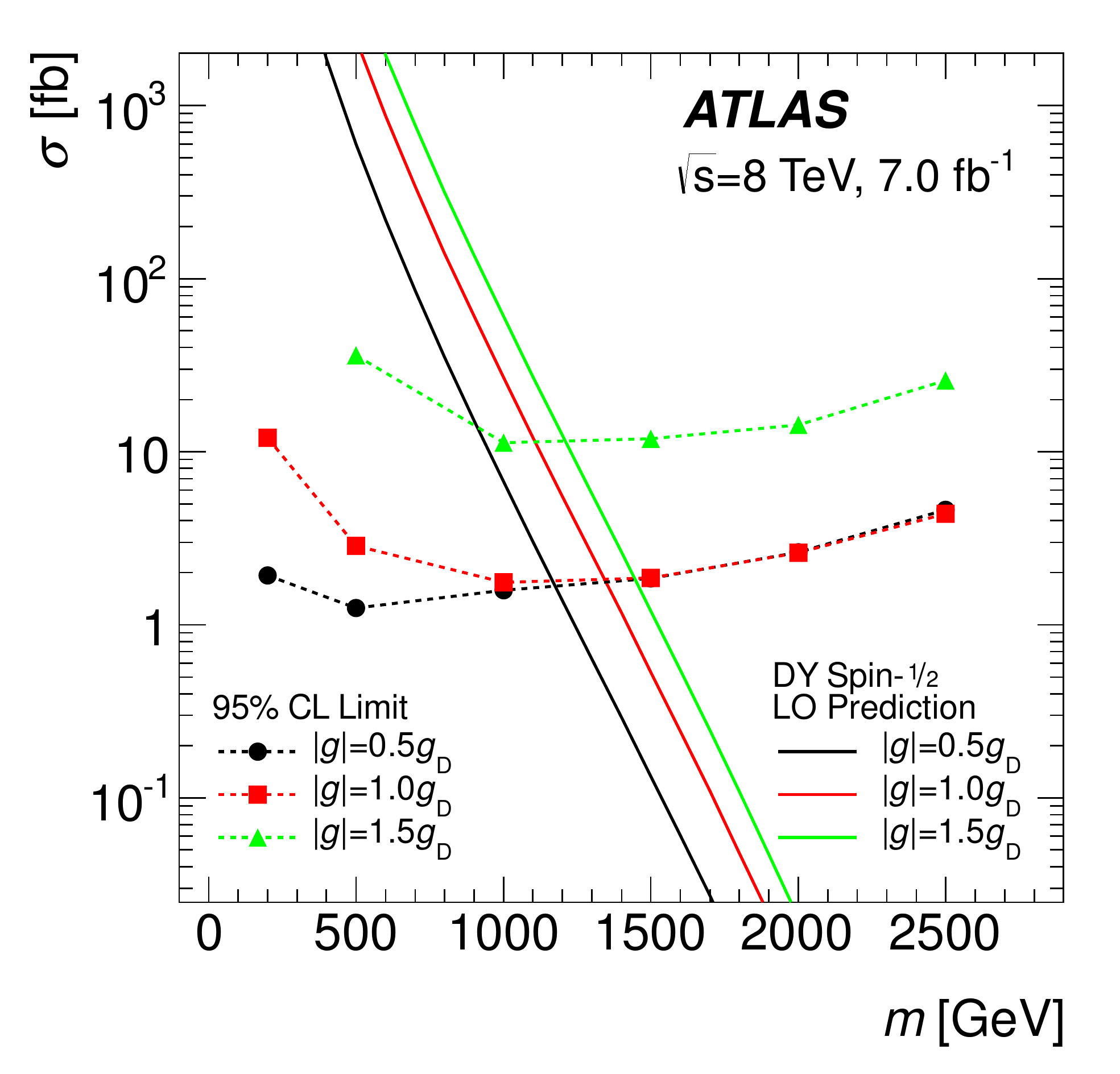}
    \includegraphics[width=0.49\linewidth]{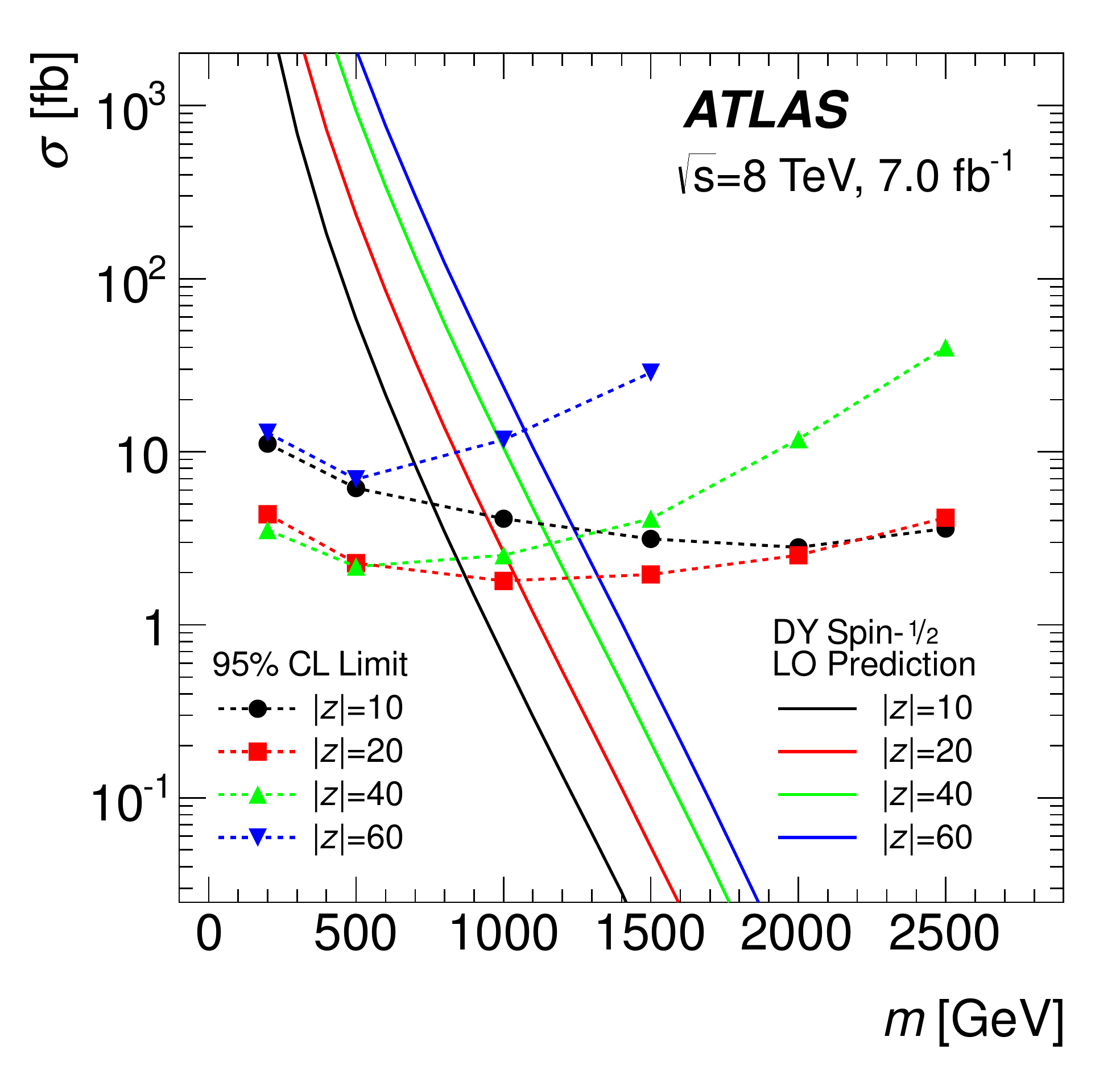}
    \includegraphics[width=0.49\linewidth]{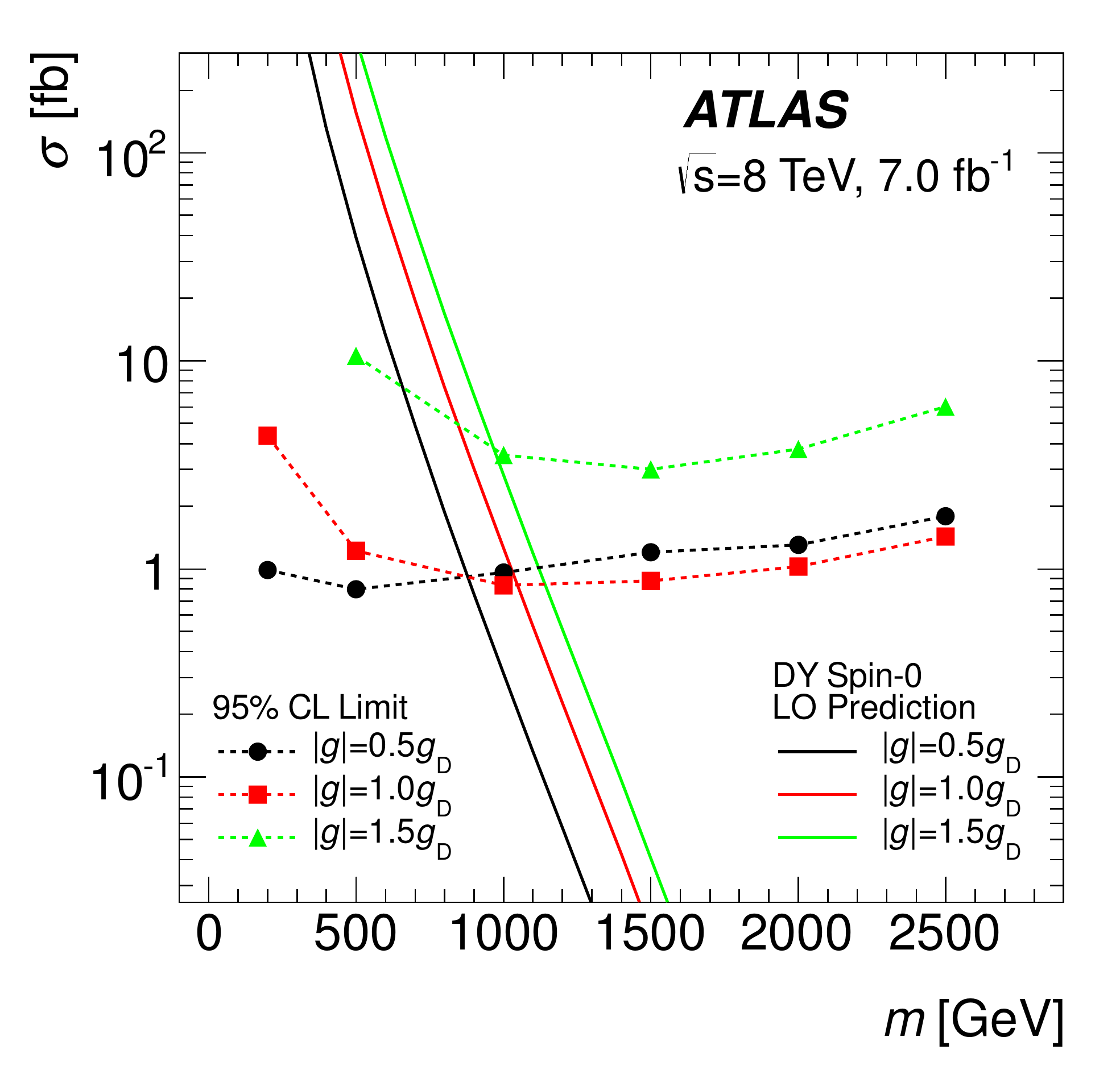}
    \includegraphics[width=0.49\linewidth]{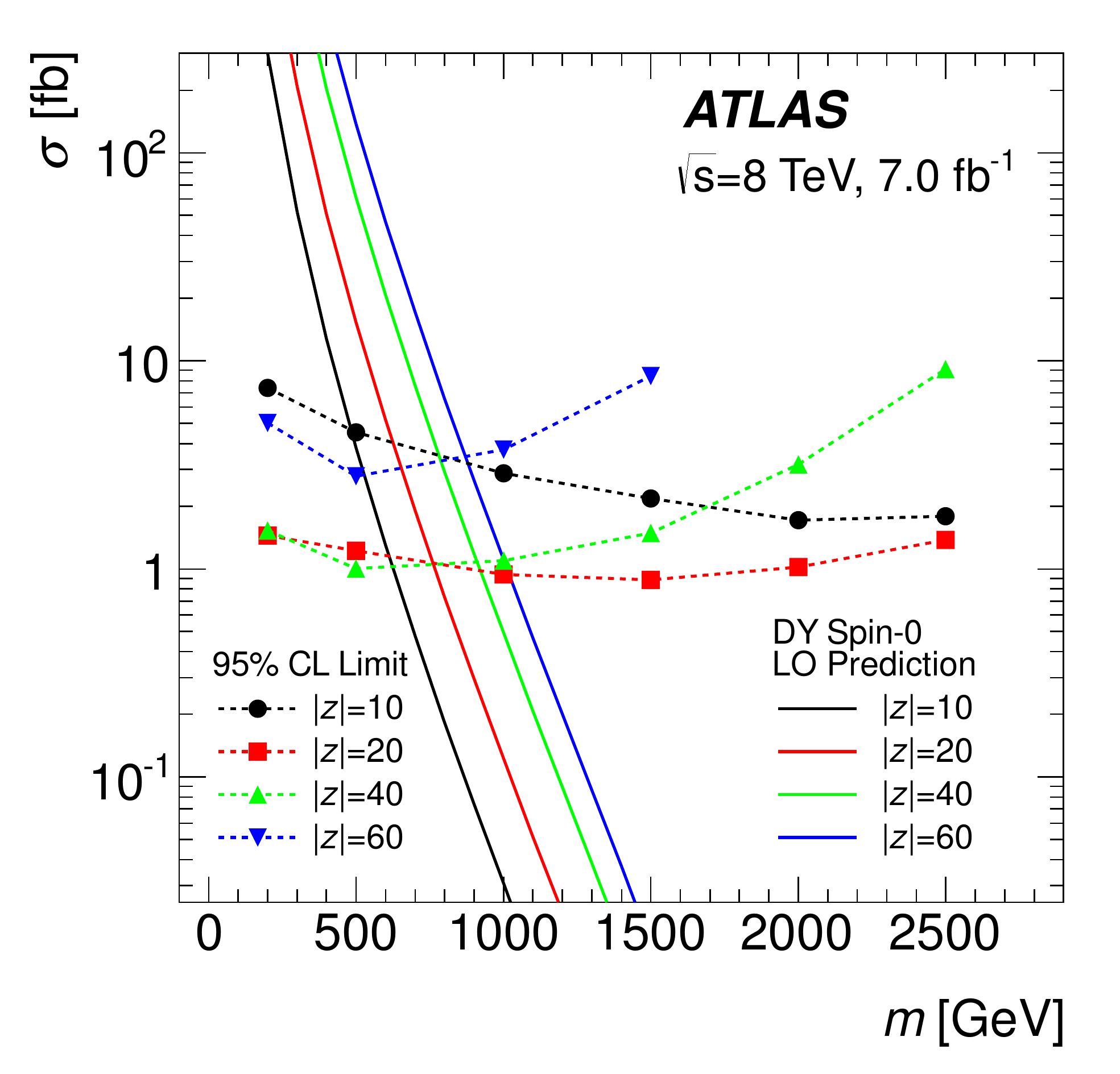}
  \end{center}
  \caption{Cross section upper limits at 95\% confidence level for DY HIP production as a
    function of HIP mass in various scenarios (dashed lines with
    markers). The upper plots are for spin-1/2 HIP production, whereas the lower plots are for spin-0 HIPs. No cross section limit is shown for mass/charge points
    with an acceptance lower than 1\%. Overlaid on the plots are the
    leading-order (LO) cross sections (solid lines). }
  \label{fig:xs}
\end{figure}

Cross section limits are driven by the selection efficiencies (and their uncertainties) for the various signal hypotheses. They are determined using the full CL$_s$ frequentist method~\cite{Read2002} for each of the HIP masses and charges.
In the fiducial regions, a 90\% signal efficiency is used (this comes from the fiducial region definition, see Section~\ref{sec:eff_fiducial} and Fig.~\ref{fig:fiducial}).
The 95\% confidence-level cross section upper limit for the fiducial regions is 0.5~fb. 
The cross section limits for DY pair-production are shown graphically as functions of mass in Fig.~\ref{fig:xs}. 

\subsection{Model-dependent mass limits}
\label{sec:mass}

As has often been pointed out in the literature (e.g., in Refs.~\cite{Fairbairn2007,DeRoeck2012a}), the accuracy of HIP mass limits is questionable due to the nonperturbative nature of the underlying process, which renders cross section predictions unreliable. However, such limits are still useful for comparing results from different searches that make similar theoretical assumptions.
In Table~\ref{tab:masslimits}, mass lower limits at 95\% confidence level are shown, obtained assuming DY production kinematic distributions for spin-1/2 and spin-0 HIPs.

\begin{table}
\caption{Lower mass limits (in GeV) at 95\% confidence level in models of spin-1/2 (top) and spin-0 (bottom) DY HIP pair production. These limits are based upon leading-order models, since the production mechanism is of a highly nonperturbative nature.}
\label{tab:masslimits}
\begin{center}
\begin{tabular}{|l|c|c|c|c|c|c|c|}
\hline
& \multicolumn{7}{c|}{Drell-Yan Lower Mass Limits [GeV]}\\
\hline
      & $|g|=0.5g_{\rm D}$ & $|g|=1.0g_{\rm D}$ & $|g|=1.5g_{\rm D}$ &  $|z|=10$ & $|z|=20$ & $|z|=40$ & $|z|=60$ \\
\hline
spin-1/2  & 1180    & 1340     & 1210    & 780      & 1050   & 1160    & 1070   \\
spin-0    & 890     & 1050     & 970     & 490      & 780    & 920     & 880   \\
\hline
\end{tabular}
\end{center}
\end{table}

\section{Conclusion}
\label{sec:conclusion}

A search for magnetic monopoles and exotic stable particles with high electric charge was performed with the ATLAS detector at the LHC using 7.0~fb$^{-1}$ of 8~TeV $pp$ collision data using a signature of a highly ionizing particle stopping in the EM calorimeter. Candidates were selected by exploiting the measured ionization in the TRT detector and the shape of the energy deposition in the EM calorimeter. No events were observed in data in the signal region. Upper limits on the production cross section were set for mass and charge points to which the search proves sensitive. A model-independent upper limit on the production cross section of 0.5~fb was obtained for signal particles with magnetic charge in the range \mbox{$0.5 g_{\rm D}\leq |g|\leq 2.0 g_{\rm D}$} and electric charge in the range $20\leq |z|\leq 60$ with masses between 200~GeV and 2500~GeV. This result is valid in well-defined fiducial regions of high and uniform event selection efficiency. Assuming Drell-Yan pair production of spin-1/2 and spin-0 charged massive particles, upper limits on the production cross section were obtained for $0.5 g_{\rm D}\leq |g|\leq 1.5 g_{\rm D}$ and $10\leq |z|\leq 60$ and masses up to 2500~GeV.

These results improve the upper limits on the production cross section for HIPs in mass and charge regions accessible to preceding experiments, and extend the limits to masses higher than 1500~GeV. Monopoles with a magnetic charge higher than $|g|=1.0g_{\rm D}$  (up to $|g|=2.0g_{\rm D}$) and exotic stable particles with an electric charge higher than $|z|=17$ (up to $|z|=60$) were probed for the first time at the LHC.

\section*{Acknowledgments}

We thank CERN for the very successful operation of the LHC, as well as the
support staff from our institutions without whom ATLAS could not be
operated efficiently.  We thank R.~Koniuk for useful contributions.

We acknowledge the support of ANPCyT, Argentina; YerPhI, Armenia; ARC, Australia; BMWFW and FWF, Austria; ANAS, Azerbaijan; SSTC, Belarus; CNPq and FAPESP, Brazil; NSERC, NRC and CFI, Canada; CERN; CONICYT, Chile; CAS, MOST and NSFC, China; COLCIENCIAS, Colombia; MSMT CR, MPO CR and VSC CR, Czech Republic; DNRF and DNSRC, Denmark; IN2P3-CNRS, CEA-DSM/IRFU, France; GNSF, Georgia; BMBF, HGF, and MPG, Germany; GSRT, Greece; RGC, Hong Kong SAR, China; ISF, I-CORE and Benoziyo Center, Israel; INFN, Italy; MEXT and JSPS, Japan; CNRST, Morocco; FOM and NWO, Netherlands; RCN, Norway; MNiSW and NCN, Poland; FCT, Portugal; MNE/IFA, Romania; MES of Russia and NRC KI, Russian Federation; JINR; MESTD, Serbia; MSSR, Slovakia; ARRS and MIZ\v{S}, Slovenia; DST/NRF, South Africa; MINECO, Spain; SRC and Wallenberg Foundation, Sweden; SERI, SNSF and Cantons of Bern and Geneva, Switzerland; MOST, Taiwan; TAEK, Turkey; STFC, United Kingdom; DOE and NSF, United States of America. In addition, individual groups and members have received support from BCKDF, the Canada Council, CANARIE, CRC, Compute Canada, FQRNT, and the Ontario Innovation Trust, Canada; EPLANET, ERC, FP7, Horizon 2020 and Marie Sk{\l}odowska-Curie Actions, European Union; Investissements d'Avenir Labex and Idex, ANR, R{\'e}gion Auvergne and Fondation Partager le Savoir, France; DFG and AvH Foundation, Germany; Herakleitos, Thales and Aristeia programmes co-financed by EU-ESF and the Greek NSRF; BSF, GIF and Minerva, Israel; BRF, Norway; the Royal Society and Leverhulme Trust, United Kingdom.

The crucial computing support from all WLCG partners is acknowledged
gratefully, in particular from CERN and the ATLAS Tier-1 facilities at
TRIUMF (Canada), NDGF (Denmark, Norway, Sweden), CC-IN2P3 (France),
KIT/GridKA (Germany), INFN-CNAF (Italy), NL-T1 (Netherlands), PIC (Spain),
ASGC (Taiwan), RAL (UK) and BNL (USA) and in the Tier-2 facilities
worldwide.
\bibliographystyle{atlasBibStyleWoTitle}

\providecommand{\href}[2]{#2}\begingroup\raggedright\endgroup


\newpage
\input{atlas_authlist}

\end{document}

%% file: atlas_authlist.tex
\begin{flushleft}
{\Large The ATLAS Collaboration}

\bigskip

G.~Aad$^{\rm 85}$,
B.~Abbott$^{\rm 113}$,
J.~Abdallah$^{\rm 151}$,
O.~Abdinov$^{\rm 11}$,
R.~Aben$^{\rm 107}$,
M.~Abolins$^{\rm 90}$,
O.S.~AbouZeid$^{\rm 158}$,
H.~Abramowicz$^{\rm 153}$,
H.~Abreu$^{\rm 152}$,
R.~Abreu$^{\rm 116}$
% ...
\bigskip
\\
% ...
$^{11}$ Institute of Physics, Azerbaijan Academy of Sciences, Baku, Azerbaijan\\
% ...
$^{85}$ CPPM, Aix-Marseille Universit{\'e} and CNRS/IN2P3, Marseille, France\\
% ...
$^{90}$ Department of Physics and Astronomy, Michigan State University, East Lansing MI, United States of America\\
% ...
$^{107}$ Nikhef National Institute for Subatomic Physics and University of Amsterdam, Amsterdam, Netherlands\\
% ...
$^{113}$ Homer L. Dodge Department of Physics and Astronomy, University of Oklahoma, Norman OK, United States of America\\
% ...
$^{116}$ Center for High Energy Physics, University of Oregon, Eugene OR, United States of America\\
% ...
$^{151}$ Institute of Physics, Academia Sinica, Taipei, Taiwan\\
$^{152}$ Department of Physics, Technion: Israel Institute of Technology, Haifa, Israel\\
$^{153}$ Raymond and Beverly Sackler School of Physics and Astronomy, Tel Aviv University, Tel Aviv, Israel\\
% ...
$^{158}$ Department of Physics, University of Toronto, Toronto ON, Canada\\
% ...
\end{flushleft}


%% file: ATLASHIP14.bbl
\providecommand{\href}[2]{#2}\begingroup\raggedright\begin{thebibliography}{}

\end{thebibliography}\endgroup


\begin{thebibliography}{10}

\bibitem{Fairbairn2007}
M.~Fairbairn, A.~Kraan, D.~Milstead, T.~Sjostrand, P.~Skands,  and T.~Sloan,
  \href{http://dx.doi.org/10.1016/j.physrep.2006.10.002}{Phys. Rept. {\bfseries
  438} (2007) 1}.

\bibitem{MoEDAL2014}
{{MoEDAL}} Collaboration, B.~Acharya {et~al.},
  \href{http://dx.doi.org/10.1142/S0217751X14300506}{Int. J. Mod. Phys. A
  {\bfseries 29} (2014) 1430050}.

\bibitem{Dirac1931}
P.~Dirac, \href{http://dx.doi.org/10.1098/rspa.1931.0130}{Proc. Roy. Soc.
  {\bfseries A 133} (1931) 60}.

\bibitem{Dirac1948}
P.~Dirac, \href{http://dx.doi.org/10.1103/PhysRev.74.817}{Phys. Rev. {\bfseries
  74} (1948) 817}.

\bibitem{Polyakov1974}
A.~Polyakov, ZhETF Pis. Red. {\bf 20}, 430 (1974). [\href{http://www.jetpletters.ac.ru/ps/1789/article_27297.pdf}{JETP Lett. {\bfseries 20} (1974) 194.}]

\bibitem{tHooft1974}
G.~'t~Hooft, \href{http://dx.doi.org/10.1016/0550-3213(74)90486-6}{Nucl. Phys.
  B {\bfseries 79} (1974) 276}.

\bibitem{Cho1997}
Y.~Cho and D.~Maison,
  \href{http://dx.doi.org/10.1016/S0370-2693(96)01492-X}{Phys. Lett. B
  {\bfseries 391} (1997) 360}.

\bibitem{MACRO2002a}
{{MACRO}} Collaboration, M.~Ambrosio {et~al.},
  \href{http://dx.doi.org/10.1140/epjc/s2002-01046-9}{Eur. Phys. J. C
  {\bfseries 25} (2002) 511}.

\bibitem{SLIM2008a}
{{SLIM}} Collaboration, S.~Balestra {et~al.},
  \href{http://dx.doi.org/10.1140/epjc/s10052-008-0597-3}{Eur. Phys. J. C
  {\bfseries 55} (2008) 57}.

\bibitem{RICE2008}
D.~P. Hogan, D.~Z. Besson, J.~P. Ralston, I.~Kravchenko,  and D.~Seckel,
  \href{http://dx.doi.org/10.1103/PhysRevD.78.075031}{Phys. Rev. D {\bfseries
  78} (2008) 075031}.

\bibitem{ANITA2011}
{{ANITA-II}} Collaboration, M.~Detrixhe {et~al.},
  \href{http://dx.doi.org/10.1103/PhysRevD.83.023513}{Phys. Rev. D {\bfseries
  83} (2011) 023513}.

\bibitem{ANTARES2012}
{{ANTARES}} Collaboration, S.~Adri\'{a}n-Martinez {et~al.},
  \href{http://dx.doi.org/10.1016/j.astropartphys.2012.02.007}{Astropart. Phys. {\bfseries  35} (2012) 634}.

\bibitem{IceCube2013}
{{IceCube}} Collaboration, R.~Abbasi {et~al.},
  \href{http://dx.doi.org/10.1103/PhysRevD.87.022001}{Phys. Rev. D {\bfseries
  87} (2013) 022001}.

\bibitem{IceCube2014}
{{IceCube}} Collaboration, M.~G. Aartsen {et~al.},
  \href{http://dx.doi.org/10.1140/epjc/s10052-014-2938-8}{Eur. Phys. J. C
  {\bfseries 74} (2014) 2938}.

\bibitem{Ross1973}
R.~Ross, P.~Eberhard, L.~Alvarez,  and R.~Watt,
  \href{http://dx.doi.org/10.1103/PhysRevD.8.698}{Phys. Rev. D {\bfseries 8}
  (1973) 698}.

\bibitem{Kovalik1986}
J.~Kovalik and J.~Kirschvink,
  \href{http://dx.doi.org/10.1103/PhysRevA.33.1183}{Phys. Rev. A {\bfseries 33}
  (1986) 1183}.

\bibitem{Jeon1995}
H.~Jeon and M.~Longo,
  \href{http://dx.doi.org/10.1103/PhysRevLett.75.1443}{Phys. Rev. Lett.
  {\bfseries 75} (1995) 1443}.

\bibitem{Bendtz2013a}
K.~Bendtz, D.~Milstead, H.-P. Hachler, A.~Hirt, P.~Mermod, P.~Michael,
  T.~Sloan, C.~Tegner,  and S.~Thorarinsson,
  \href{http://dx.doi.org/10.1103/PhysRevLett.110.121803}{Phys. Rev. Lett.
  {\bfseries 110} (2013) 121803}.

\bibitem{Aubert:1982zi}
B.~Aubert, P.~Musset, M.~Price,  and J.~Vialle,
\href{http://dx.doi.org/10.1016/0370-2693(83)90488-4}{Phys. Lett. B {\bfseries
  120} (1983) 465}.

\bibitem{Musset:1983ii}
P.~Musset, M.~Price,  and E.~Lohrmann,
\href{http://dx.doi.org/10.1016/0370-2693(83)90270-8}{Phys. Lett. B {\bfseries
  128} (1983) 333}.

\bibitem{D01998}
{{D0}} Collaboration, B.~Abbott {et~al.},
  \href{http://dx.doi.org/10.1103/PhysRevLett.81.524}{Phys. Rev. Lett.
  {\bfseries 81} (1998) 524}.

\bibitem{Aktas:2004qd}
{{H1}} Collaboration, A.~Aktas {et~al.},
\href{http://dx.doi.org/10.1140/epjc/s2005-02201-6}{Eur. Phys. J. C {\bfseries
  41} (2005) 133}.

\bibitem{CDF2006}
{{CDF}} Collaboration, A.~Abulencia {et~al.},
  \href{http://dx.doi.org/10.1103/PhysRevLett.96.201801}{Phys. Rev. Lett.
  {\bfseries 96} (2006) 201801}.

\bibitem{OPAL2008}
{{OPAL}} Collaboration, G.~Abbiendi {et~al.},
  \href{http://dx.doi.org/10.1016/j.physletb.2008.03.057}{Phys. Lett. B
  {\bfseries 663} (2008) 37}.

\bibitem{ATLAS2012a}
{{ATLAS}} Collaboration, G.~Aad {et~al.},
  \href{http://dx.doi.org/10.1103/PhysRevLett.109.261803}{Phys. Rev. Lett.
  {\bfseries 109} (2012) 261803}.

\bibitem{DeRoeck2012b}
A.~De~Roeck, H.-P. {H\"achler}, A.~M. Hirt, M.~Dam~Joergensen, A.~Katre,
  P.~Mermod, D.~Milstead,  and T.~Sloan,
  \href{http://dx.doi.org/10.1140/epjc/s10052-012-2212-x}{Eur. Phys. J. C
  {\bfseries 72} (2012) 2212}.

\bibitem{Bendtz2014} K.~Bendtz, A.~Katre, D.~Lacarr{\`e}re, P.~Mermod,
  D.~Milstead, J.~Pinfold, and R.~Soluk, in
  \href{http://dx.doi.org/10.1142/9789814603164_0059}{Proceedings of
    the 14th ICATPP Conference, Astroparticle, Particle, Space Physics
    and Detectors for Physics Applications} (World Scientific,
  Singapore, 2014) p. 382.

\bibitem{DeRoeck2012a}
A.~De~Roeck, A.~Katre, P.~Mermod, D.~Milstead,  and T.~Sloan,
  \href{http://dx.doi.org/10.1140/epjc/s10052-012-1985-2}{Eur. Phys. J. C
  {\bfseries 72} (2012) 1985}.

\bibitem{ATLAS2011a}
{{ATLAS}} Collaboration, G.~Aad {et~al.},
  \href{http://dx.doi.org/10.1016/j.physletb.2011.03.033}{Phys. Lett. B
  {\bfseries 698} (2011) 353}.

\bibitem{ATLAS2015}
{ATLAS Collaboration},
  \href{http://dx.doi.org/10.1140/epjc/s10052-015-3534-2}{Eur. Phys. J. C
  {\bfseries 75} (2015) 362}.

\bibitem{CMS2013a}
{{CMS}} Collaboration, S.~Chatrchyan {et~al.},
  \href{http://dx.doi.org/10.1007/JHEP07(2013)122}{JHEP {\bfseries 07} (2013)
  122}.

\bibitem{ATLAS2008}
{ATLAS Collaboration},
  \href{http://dx.doi.org/10.1088/1748-0221/3/08/S08003}{JINST {\bfseries 3}
  (2008) S08003}.

\bibitem{ATLAS2010b}
{ATLAS Collaboration},
  \href{http://dx.doi.org/10.1140/epjc/s10052-010-1354-y}{Eur. Phys. J. C
  {\bfseries 70} (2010) 723}.

\bibitem{MadGraph}
J.~Alwall, M.~Herquet, F.~Maltoni, O.~Mattelaer,  and T.~Stelzer,
  \href{http://dx.doi.org/10.1007/JHEP06(2011)128}{JHEP {\bfseries 1106} (2011)
  128}.

\bibitem{CTEQ6L1}
J.~Pumplin, D.~R. Stump, J.~Huston, H.-L. Lai, P.~Nadolsky,  and W.-K. Tung,
  \href{http://dx.doi.org/10.1088/1126-6708/2002/07/012}{JHEP {\bfseries 07}
  (2002) 012}.

\bibitem{Pythia6}
T.~Sj{\"o}strand, S.~Mrenna,  and P.~Skands,
  \href{http://dx.doi.org/10.1088/1126-6708/2006/05/026}{JHEP {\bfseries 05}
  (2006) 026}.

\bibitem{Pythia8}
T.~Sj{\"o}strand, S.~Mrenna,  and P.~Skands,
  \href{http://dx.doi.org/10.1016/j.cpc.2008.01.036}{Comput. Phys. Comm.
  {\bfseries 178} (2008) 852}.

\bibitem{Ahlen1975}
S.~Ahlen, \href{http://dx.doi.org/10.1103/PhysRevD.14.2935}{Phys. Rev. D
  {\bfseries 14} (1976) 2935}.

\bibitem{Ahlen1978}
S.~Ahlen, \href{http://dx.doi.org/10.1103/PhysRevD.17.229}{Phys. Rev. D
  {\bfseries 17} (1978) 229}.

\bibitem{ATLAS2010c}
{ATLAS Collaboration},
  \href{http://dx.doi.org/10.1140/epjc/s10052-010-1429-9}{Eur. Phys. J. C
  {\bfseries 70} (2010) 823}.

\bibitem{GEANT42003}
{{GEANT4}} Collaboration, S.~Agostinelli {et~al.},
  \href{http://dx.doi.org/10.1016/S0168-9002(03)01368-8}{Nucl. Instrum. Meth. A
  {\bfseries 506} (2003) 250}.

\bibitem{Burdin2012}
S.~Burdin, M.~Horbatsch,  and W.~Taylor,
  \href{http://dx.doi.org/10.1016/j.nima.2011.10.044}{Nucl. Inst. Meth. A
  {\bfseries 664} (2012) 111}.

\bibitem{POWHEG}
S.~Alioli, P.~Nason, C.~Oleari,  and E.~Re,
\href{http://dx.doi.org/10.1088/1126-6708/2008/07/060}{JHEP {\bfseries 07}
  (2008) 060}.

\bibitem{ATLAS2012f}
{ATLAS Collaboration}, ATL-PHYS-PUB-2012-003 (2012),
  {\url{http://cds.cern.ch/record/1474107}}.

\bibitem{ATLAS2012d}
{ATLAS Collaboration}, ATLAS-CONF-2012-048 (2012),
  {\url{http://cds.cern.ch/record/1450089}}.

\bibitem{Lampl2008}
W.~Lampl, S.~Laplace, D.~Lelas, P.~Loch, H.~Ma, S.~Menke, S.~Rajagopalan,
  D.~Rousseau, S.~Snyder,  and G.~Unal, ATL-LARG-PUB-2008-002 (2008),
  {\url{http://cds.cern.ch/record/1099735}}.

\bibitem{ATLAS2012e}
{{ATLAS}} Collaboration, G.~Aad {et~al.},
  \href{http://dx.doi.org/10.1088/1748-0221/7/01/P01013}{JINST {\bfseries 7}
  (2012) P01013}.

\bibitem{ATLASLUMI2013}
{ATLAS Collaboration},
  \href{http://dx.doi.org/10.1140/epjc/s10052-013-2518-3}{Eur. Phys. J. C
  {\bfseries 73} (2013) 2518}.

\bibitem{Read2002}
A.~Read, \href{http://dx.doi.org/10.1088/0954-3899/28/10/313}{J. Phys. G
  {\bfseries 28} (2002) 2693}.

\end{thebibliography}
